\documentclass[twocolumn, switch]{article} 

\usepackage{preprint}

\usepackage{amsmath, amsthm, amssymb, amsfonts}

\newcounter{ass}                           
\newcounter{rem}
\newcounter{lem}

\newcounter{def}
\newcounter{pro}
\newcounter{thm}

\newtheorem{assumption}[ass]{Assumption}
\newtheorem{remark}[rem]{Remark}
\newtheorem{lemma}[lem]{Lemma}

\newtheorem{definition}[def]{Definition}
\newtheorem{proposition}[pro]{Proposition}
\newtheorem{theorem}[thm]{Theorem}

\usepackage{appendix}

\usepackage[numbers,square]{natbib}
\usepackage[utf8]{inputenc}	
\usepackage[T1]{fontenc}	
\usepackage{xcolor}		
\usepackage[colorlinks = true,
            linkcolor = purple,
            urlcolor  = blue,
            citecolor = cyan,
            anchorcolor = black]{hyperref}	
\usepackage{booktabs} 		
\usepackage{nicefrac}		
\usepackage{microtype}		
\usepackage{lineno}		
\usepackage{float}			
\usepackage{multirow}
\usepackage{graphicx}
\graphicspath{ {./images/} }

\usepackage{lipsum}		

\usepackage{newfloat}
\DeclareFloatingEnvironment[name={Supplementary Figure}]{suppfigure}
\usepackage{sidecap}
\sidecaptionvpos{figure}{c}

\usepackage{titlesec}
\titlespacing\section{0pt}{12pt plus 3pt minus 3pt}{1pt plus 1pt minus 1pt}
\titlespacing\subsection{0pt}{10pt plus 3pt minus 3pt}{1pt plus 1pt minus 1pt}
\titlespacing\subsubsection{0pt}{8pt plus 3pt minus 3pt}{1pt plus 1pt minus 1pt}

\usepackage{colortbl}
\usepackage{hhline}
\definecolor{myblue}{RGB}{0,102,204}

\usepackage{tikz,xcolor,hyperref}

\definecolor{lime}{HTML}{A6CE39}
\DeclareRobustCommand{\orcidicon}{
	\begin{tikzpicture}
	\draw[lime, fill=lime] (0,0) 
	circle [radius=0.16] 
	node[white] {{\fontfamily{qag}\selectfont \tiny ID}};
	\draw[white, fill=white] (-0.0625,0.095) 
	circle [radius=0.007];
	\end{tikzpicture}
	\hspace{-2mm}
}
\foreach \x in {A, ..., Z}{\expandafter\xdef\csname orcid\x\endcsname{\noexpand\href{https://orcid.org/\csname orcidauthor\x\endcsname}
			{\noexpand\orcidicon}}
}

\title{Robust contraction-based model predictive control for nonlinear systems}

\author{
	Marco Polver\\
	Dipartimento di Ingegneria Gestionale, dell'Informazione e della Produzione\\
	Università degli Studi di Bergamo (24044 Dalmine, Bergamo, Italy)\\
	\texttt{polvermarco@gmail.com}\\
	\And
	Daniel Limon\\
	Departamento de Ingeniería de Sistemas y Automática, Escuela Técnica Superior de Ingeniería\\Universidad de Sevilla (41092 Sevilla, Spain)\\
	\texttt{dlm@us.es}\\
	\And
	Fabio Previdi\\
	Dipartimento di Ingegneria Gestionale, dell'Informazione e della Produzione\\
	Università degli Studi di Bergamo (24044 Dalmine, Bergamo, Italy)\\
	\texttt{fabio.previdi@unibg.it}\\
	\And
	Antonio Ferramosca\\
	Dipartimento di Ingegneria Gestionale, dell'Informazione e della Produzione\\
	Università degli Studi di Bergamo (24044 Dalmine, Bergamo, Italy)\\
	\texttt{antonio.ferramosca@unibg.it}\\
}

\begin{document}

\twocolumn[ 
  \begin{@twocolumnfalse} 

    \maketitle

    \begin{abstract}
      Model Predictive Control (MPC) is a widely known control method that has proved to be particularly effective in multivariable and constrained control. Closed-loop stability and recursive feasibility can be guaranteed by employing accurate models in prediction and suitable terminal ingredients, i.e. the terminal cost function and the terminal constraint. Issues might arise in case of model mismatches or perturbed systems, as the state predictions could be inaccurate, and nonlinear systems for which the computation of the terminal ingredients can result challenging. In this manuscript, we exploit the properties of component-wise uniformly continuous and stabilizable systems to introduce a robust contraction-based MPC for the regulation of nonlinear perturbed systems, that employs an easy-to-design terminal cost function, does not make use of terminal constraints, and selects the shortest prediction horizon that guarantees the stability of the closed-loop system.
    \end{abstract}
    \vspace{0.35cm}

  \end{@twocolumnfalse} 
] 



\section{Introduction}
\label{sec:introduction}
Model Predictive Control is an advanced control strategy that computes the optimal control sequence that has to be applied to the plant by solving online, at each sampling instant, a finite-horizon open-loop optimal control problem, using the current state of the plant as the initial state \cite{Camacho2007}, \cite{mayne2000constrained}. In order to close the loop, the first control action of the computed sequence is applied to the plant, the new state is measured, the prediction window is moved one step ahead and a new optimal control problem is solved starting from the new state of the system, in a receding horizon fashion. The main advantage of MPC over different control methods, e.g. PID controllers, consists in its native ability to control multiple-input multiple-output (MIMO) systems that are subject to input, state and output constraints. The stability of the closed loop and the recursive feasibility of the controller are usually ensured by design, i.e. by employing an accurate prediction model, by choosing a suitable prediction horizon and by thoughtfully designing the ``terminal ingredients'' of the MPC. These include the terminal cost function, which is usually required to be a control Lyapunov function (CLF), and the terminal constraint, namely a constraint that is imposed on the state that is predicted at the end of the prediction horizon, which is often imposed to belong to an invariant set for the system controlled by a locally feasible and stabilizing control law \cite{mayne2000constrained}, \cite{grune2017nonlinear}, \cite{Mayne2014model}.

Nevertheless, the computation of the terminal ingredients might prove difficult for some nonlinear systems \cite{grune2017nonlinear}. In particular, the locally stabilizing control law is often chosen to be a linear control law based on a local linearization of the system, and the terminal cost function is usually chosen as a quadratic cost function that employs a weight matrix that is usually also obtained starting from a local linearization of the system, e.g. the Algebraic Riccati Equation (ARE) solution. However, if the local linearization is not fully controllable and the uncontrollable states are not locally asymptotically stable, such methods fail at providing a CLF and an invariant set.
Many solutions have been proposed to simplify the design of the terminal ingredients.
For instance, several methods have been proposed for guaranteeing closed-loop stability also without terminal constraints \cite{Limon2006unconstrained}, \cite{rawlings2017model}, but they still rely on the usage of a terminal cost function that locally behaves as a CLF. Another straightforward solution consists in imposing a terminal equality constraint, which makes it possible to avoid the design of a CLF and an invariant set. However, MPCs with terminal equality constraints usually need longer prediction horizons to ensure recursive feasibility. Furthermore, an exact satisfaction of a terminal equality constraint can require an infinite number of iterations in the case of some nonlinear systems \cite{allgower1998qihmpc}. Another alternative is the so-called Lyapunov-based MPC formulation \cite{christofides2011networked}, which basically imposes explicit constraints on the employed cost function to ensure it has all the properties of a Lyapunov function. Nonetheless, such constraints complicate the optimization problem to be solved and are often based on the knowledge of another stabilizing control law.

It is also often the case that we have to deal with model uncertainties or perturbed systems, issues that can be addressed by resorting to robust MPC formulations like min-max MPC, e.g. \cite{bemporad2003min}, \cite{diehl2007formulation}, \cite{Diehl2004minmax}, \cite{kerrigan2000robust}, and \cite{lazar2006min}, and tube-based MPC, as \cite{Chisci2001_disturbances}, \cite{Mayne2011tube}, and \cite{Mayne2005_robust_MPC}.
The aim of this work is therefore to design a recursively feasible MPC for the control of perturbed nonlinear systems for which the computation of standard terminal ingredients is challenging or impossible. In detail, we here introduce an MPC that is able to stabilize the closed-loop system with relatively short prediction horizons, without stability-related terminal constraints, and using an easy-to-design terminal cost function. 

The employment of simplified terminal ingredients and short prediction horizons is made possible by a clever exploitation of the contractivity properties of stabilizable systems. Several contraction-based MPC algorithms were proposed in the past, e.g. \cite{de2000contractive}. However, these MPC formulations explicitly required the state $x(k)$ of the system to decrease in some norm by imposing the constraints
$$\Vert x(j \vert k) \Vert \leq \gamma \Vert x(k) \Vert , \; \gamma \in [0,1),$$
for $j \in [1,N_p]$, where $N_p$ is the prediction horizon of the MPC, $x(k)$ is the state of a discrete-time system at instant $k$ and $x(j \vert k)$ is the $j$-step-ahead prediction of the system state with initial condition $x(k)$. An issue of such formulations resides in the receding-horizon nature of MPC. Indeed, the contraction of the state norm might not be guaranteed in closed loop even if the aforementioned constraints are verified in open loop. Nonetheless, the contraction property allows to design MPC algorithms with short prediction horizons, which is the main reason why contraction-based MPC has raised interest again during the last few years, leading to new implementations like those presented in \cite{sencio2020terminal} and \cite{zhao2018robust}. In particular, \cite{alamir2017contraction} presents an MPC algorithm with variable prediction horizon that ensures contractivity without imposing any terminal constraints, thanks to the fact that any positive definite function $\Gamma(\cdot)$ of the system state satisfies the contraction property for an appropriate horizon when the system under control is stabilizable \cite[Chapter~4]{alamir2006stabilization}.

In this work, the concepts introduced in \cite{alamir2017contraction} about a contractive MPC without terminal constraints are integrated with the theory about input-to-state (practical) stability of systems controlled by MPC controllers, for which we refer to \cite{Limon2009iss}, to present a stabilizing robust contraction-based MPC algorithm with variable prediction horizon, that is free of terminal constraints and makes use of easy-to-design terminal cost functions.

The remainder of the manuscript is organized as follows. Section \ref{sec:preliminaries} provides some preliminary concepts about contractivity and formulates the control problem. In Section \ref{sec:robust_cmpc}, the here-proposed robust contraction-based MPC for nonlinear systems is presented. In Section \ref{sec:case_studies}, our method is validated on a nonholonomic uncertain system and an uncertain four-tank system. Conclusions are stated at the end of the manuscript.

\subsection*{Notation}
$\mathbb{R}_{> i}$ ($\mathbb{R}_{\geq i}$) denotes the set of real numbers that are greater than (or equal to) $i$.
Concatenations of column vectors are represented as $(a,b) \coloneqq  [a^\top \; b^\top]^\top$.
Bold lowercase letters denote sequences of scalars or vectors ($\mathbf{u} = \{ u(k),\hdots,u(j) \}$).
Calligraphic and Greek uppercase letters are used to indicate function classes ($\mathcal{K}$, $\mathcal{K}_\infty$ and $\mathcal{KL}$), and compact sets, e.g. $\mathcal{X}$, $\mathcal{U}$, and $\Omega$. The $j$-ary Cartesian power of a set $\mathcal{U}$, i.e. $\mathcal{U} \times \mathcal{U} \times \hdots \times \mathcal{U}$, is denoted $\mathcal{U}^j$. Unless otherwise specified, $\Vert x \Vert$ denotes the 2-norm of the vector $x$.

The solution of system $x(k+1) = f \left( x(k),u(k) \right)$, where $x$ is the system state and $u$ the system input, at sampling time $j$, starting from the initial condition $x$, given the input sequence $\mathbf{u} \coloneqq \left\{ u(0),\hdots,u(j-1) \right\}$, is denoted $\phi(j;x,\mathbf{u})$.
In the same way, the solution of system $x(k+1) = f \left( x(k),u(k),w(k) \right)$, where $w$ is a perturbation, at sampling time $j$, starting from the initial condition $x$, given the input sequence $\mathbf{u} \coloneqq \left\{ u(0),\hdots,u(j-1) \right\}$ and the perturbation signal $\mathbf{w} \coloneqq \left\{ w(0),\hdots,w(j-1) \right\}$, is denoted $\phi(j;x,\mathbf{u},\mathbf{w})$. 

A function $\alpha: \; \mathbb{R}_{\geq 0} \rightarrow \mathbb{R}_{\geq 0}$
is a $\mathcal{K}$-function if $\alpha(0) = 0$ and it is strictly increasing. 
A function $\alpha: \; \mathbb{R}_{\geq 0} \rightarrow \mathbb{R}_{\geq 0}$
is a $\mathcal{K}_\infty$-function if it is a $\mathcal{K}$-function and is unbounded.
A function $\beta: \; \mathbb{R}_{\geq 0} \times \mathbb{Z}_{\geq 0} \rightarrow \mathbb{R}_{\geq 0}$ is of class $\mathcal{KL}$ if for every fixed $k$ the function $\beta(\cdot,k) \in \mathcal{K}$ and for fixed $s$ the function $\beta(s,\cdot)$ is non-increasing and $\lim_{k \to \infty} \beta(s,k) = 0$.

\section{Preliminaries}
\label{sec:preliminaries}
In this section, we first explain the connection between the stabilizability and the contractivity of dynamic systems. Finally, we introduce the control problem for which our robust contraction-based MPC is meant.

\subsection{Contractivity of stabilizable systems}
\label{subsec:contractivity}
We here provide a definition of stabilizability for generic nonlinear discrete-time systems of the form
\begin{equation}
	\label{eq:nonlinear_system}
	x(k+1) = f \left( x(k),u(k) \right), 
\end{equation}
where $x(k) \in \mathbb{R}^n$ is the system state and $u(k) \in \mathbb{R}^m$ is the system input. The following definition is a discrete-time adaptation of the stabilizability definition provided in \cite[Chapter~4]{alamir2006stabilization}.
\begin{definition}[Stabilizability]
	Given the system \eqref{eq:nonlinear_system}, if there exists a control law $u(k) = \kappa \left( x(k) \right)$ such that the closed-loop system $x(k+1) = f_{cl} \left( x(k) \right) \coloneqq f \left( x(k), \kappa \left( x(k) \right) \right)$ admits $x = 0$ as globally asymptotically stable (GAS) equilibrium with some continuous Lyapunov function $V(x)$, i.e.
	\begin{equation}
		\label{eq:lyapunov_function}
		V \left( f_{cl}(x) \right) - V \left( x \right) \leq -\alpha_V \left( x \right),
	\end{equation}
	where $\alpha_V(\cdot)$ is a positive definite function of the state, then the system \eqref{eq:nonlinear_system} is stabilizable.
\end{definition}
The last definition simply states that a system is said to be stabilizable if there exists at least one control law that renders the origin a GAS equilibrium. It is not necessary to know the control law $\kappa(\cdot)$ and the related Lyapunov function $V(\cdot)$. 

The connection between stabilizability and contractivity is explained in \cite[Proposition~4.1]{alamir2006stabilization} and here reported for the sake of completeness.

\begin{proposition}
	If the system \eqref{eq:nonlinear_system} is stabilizable, then for all the compact subsets $\mathcal{X}_\mathrm{sub} \subset \mathbb{R}^n$ and for all the positive definite functions $\Gamma(\cdot)$, there exist a contraction factor $\gamma \in (0,1)$ and a finite $N_p \in \mathbb{N}$ such that, for all $x \in \mathcal{X}_\mathrm{sub}$, the condition
	\begin{equation}
		\label{eq:contraction}
		\min_\mathbf{u} \min_{q \in \{1,\hdots,N_p\}} \Gamma \left( \phi \left( q;x,\mathbf{u} \right) \right) \leq \gamma \Gamma(x),
	\end{equation}
	where $\mathbf{u} \coloneqq \{ u(0),\hdots,u(q-1) \}$, is verified.
\end{proposition}
As will become clear in the following sections, this connection between stabilizability and contractivity can turn out to be useful in the design of predictive controllers with short prediction horizons and without terminal constraints.

\subsection{Problem formulation}
\label{subsec:problem_formulation}
We consider a perturbed nonlinear discrete-time time-invariant system described by the equation
\begin{equation}
	\label{eq:system_equation}
	\begin{split}
		x(k+1) &= f\big(x(k),u(k),w(k)\big)
	\end{split}
\end{equation}
where $x(k) \in \mathbb{R}^n$ is the system state, $u(k) \in \mathbb{R}^m$ is the controlled input, and $w(k) \in \mathcal{W} \subset \mathbb{R}^r$ is an unknown but bounded perturbation that lies in a known hypercube $\mathcal{W} \coloneqq \left\{ w: \; \vert w_i \vert \leq \bar{w}_i, \; \forall i \in [1,r] \right\}$. The system state $x$ is assumed to be measured at each sampling instant $k \geq 0$.

The system is subject to hard state and input constraints, namely
\begin{equation}
	\label{eq:system_constraints}
	x(k) \in \mathcal{X}, \; u(k) \in \mathcal{U}, \quad \forall k \geq 0.
\end{equation}
In order to ensure the recursive feasibility and stability of the proposed robust contraction-based MPC, the state function $f(\cdot,\cdot,\cdot)$ and the sets $\mathcal{X}$ and $\mathcal{U}$ must meet the conditions provided by the following assumptions.
\begin{assumption}[Continuity of the state function]
	\label{ass:cmpc_continuity}
	The state function $f(\cdot,\cdot,\cdot)$ is such that each component $f_i(\cdot,\cdot,\cdot)$, $\forall i \in [1,n]$, is component-wise uniformly continuous. Consequently, there exist functions $\sigma_{x,ia}(\cdot),\sigma_{u,ib}(\cdot),\sigma_{w,ic}(\cdot) \in \mathcal{K}$ such that, for all $i \in [1,n]$,
	\begin{equation}
		\label{eq:cmpc_componentwise_uniform_continuity}
		\begin{split}
			\vert f_i(x,u,w) \!-\! f_i(\check{x},\check{u},\check{w}) \vert &\leq \sum_{a = 1}^n \sigma_{x,ia} \left( x_a \!-\! \check{x}_a \right) + \sum_{b = 1}^m \sigma_{u,ib} \left( \vert u_b \!-\! \check{u}_b \vert \right)\\ &+ \sum_{c = 1}^r \sigma_{w,ic} \left( \vert w_c \!-\! \check{w}_c \vert \right),
		\end{split}
	\end{equation}
	for all $(x,u,w)$ and $(\check{x},\check{u},\check{w})$ in $\mathcal{X} \times \mathcal{U} \times \mathcal{W}$.
\end{assumption} 
\begin{assumption}[Properties of the constraint sets]
	\label{ass:cmpc_constraint_set_properties}
	The sets $\mathcal{X}$ and $\mathcal{U}$ are compact and their interior is non-empty.
\end{assumption}
\begin{remark}
	Within the family of component-wise uniformly continuous functions, we find also component-wise Lipschitz continuous functions, which is a wide family of functions. 
	
	Indeed, as the domain $\mathcal{X} \times \mathcal{U} \times \mathcal{W}$ is compact, Lipschitz continuity is a condition that is easy to verify. Moreover, as proved in \cite{Manzano2019_Choki}, Lipschitz continuity implies also component-wise Lipschitz continuity.
\end{remark}
Given that the robust MPC for regulation discussed here is based on the contractivity of the system, we provide a contractivity assumption.
\begin{assumption}[Contractivity of the nominal system]
	\label{ass:cmpc_contractivity}
	There exist a positive definite function $\Gamma(x)$, a contraction factor $\gamma \in (0,1)$, and a prediction horizon $N_p \in \mathbb{N} \setminus \{0\}$ such that, for any initial conditions $x \in \mathcal{X}$, there exists a control sequence $\mathbf{u} \in \mathcal{U}^{N_p}$ that verifies:
	\begin{equation}
		\label{eq:cmpc_contractivity}
		\underline{\Gamma}(x,\mathbf{u},N_p) \coloneqq \min_{j=1}^{N_p} \Gamma \left( \phi (j;x,\mathbf{u},\mathbf{0}) \right) \leq \gamma \Gamma(x).
	\end{equation}
	Moreover, the contractive function $\Gamma(x)$ is uniformly continuous in $\mathcal{X}$, i.e. there exists a $\mathcal{K}$ function $\sigma_\Gamma(\cdot)$ such that
	\begin{equation}
		\label{eq:cmpc_Gamma_uniform_continuity}
		\left\vert \Gamma(x) - \Gamma(\check{x}) \right\vert \leq \sigma_\Gamma \left( \left\Vert x - \check{x} \right\Vert \right), \; \forall x,\check{x} \in \mathcal{X}.
	\end{equation}
	Finally, the function $\Gamma(\cdot)$ is upper-bounded by a $\mathcal{K}_\infty$ function $\alpha_\Gamma(\cdot)$, such that
	\begin{equation}
		\label{eq:cmpc_Gamma_upper_bound}
		\Gamma(x) \leq \alpha_\Gamma(\Vert x \Vert), \; \forall x \in \mathcal{X}.
	\end{equation}
\end{assumption}
Henceforward, the minimizer of \eqref{eq:cmpc_contractivity} over an horizon of length $N_p$ and given the initial condition $x$ is denoted $j_\mathrm{opt}(x, \mathbf{u}, N_p)$.

Note that Assumption \ref{ass:cmpc_contractivity} does not require all the trajectories providing the best contraction factor to lie inside $\mathcal{X}$, which was instead the case of \cite{alamir2017contraction}. This is because \cite{alamir2017contraction} requires the exact knowledge of the compact set that ensures the desired contraction, which coincides with the domain of attraction of their contraction-based MPC. 
As the computation of such set is challenging, we do not require to analytically find the region of attraction of our MPC and we rely on recursive feasibility to ensure that, if the initial condition renders the optimal control problem feasible, feasibility will be guaranteed also afterwards.
\begin{remark}
	The choice of the contractive function $\Gamma(\cdot)$ and the computation of $\gamma$ can be carried out in different ways. In the case study presented in \cite{alamir2017contraction}, an analytical way of finding $\gamma$ for a given contractive function and a particular system is reported. However, also other methods can be considered. For example, given that $\Gamma(x)$ is only required to be positive definite and uniformly continuous on a compact set, several candidates can be tested offline to find the one ensuring the desired contraction within a short horizon, and the contraction factor $\gamma$ can be estimated by means of Monte Carlo simulations or by solving unconstrained optimal control problems with the aim of minimizing the value of the contractive function within the horizon $[1,N_p]$, starting from different initial conditions $x_0 \in \mathcal{X}$.
\end{remark}
The aim of this work is to design an MPC that can drive the system state to an equilibrium $x_s = f \left( x_s, u_s,0 \right)$. Without loss of generality, in the next sections we assume to regulate the system state to the origin.

\section{The proposed robust contraction-based MPC}
\label{sec:robust_cmpc}
In this section, we first delve into the further properties that the system under control must have to enable the correct design of our robust contraction-based MPC. Then, we present two formulations of the proposed MPC, proving the stability and feasibility properties of both.

\subsection{Robust design of the controller}
\label{subsec:cmpc_robust_design}
Our MPC belongs to the family of open-loop tube-based predictive controllers, hence its recursive feasibility is guaranteed by imposing that the nominal system trajectory stays inside a sequence of tightened constraints \cite{Limon2009iss,Limon2010_robust_NMPC}, and we suppose not to exploit any feedback control laws to reduce the effect of the perturbations on the state predictions.
\begin{remark}
	It is still possible to employ the theoretical concepts introduced in this manuscript in feedback MPC schemes that require to find an optimal sequence of parameterized control laws. The main difference in such case is that a contractive function $\Gamma(\cdot)$ should be found for the nominal closed-loop system under the chosen feedback control law. Then, the presented contraction-based MPC should be modified in order to have the free parameters of the chosen control law as decision variables.
\end{remark}
The computation of the needed tightened constraints requires the existence of the sequences $\left\{ \mathcal{F}(j) \right\}_{j \geq 0}$ and $\left\{ \mathcal{R}(j) \right\}_{j \geq 0}$, whose properties are outlined in the following two assumptions.
\begin{assumption}[Bounds on the effect of the initial condition on the state trajectory]
	\label{ass:cmpc_F_sequence}
	The sequence $\{ \mathcal{F}(j) \}_{j \geq 0}$ is such that:
	\begin{itemize}
		\item $\mathcal{F}(0) \coloneqq \{ x \in \mathbb{R}^n: \vert x_i \vert \leq \sum_{a=1}^r \sigma_{w,ia} (\bar{w}_a), \forall i \in [1,n] \}$;
		\item $\forall j > 0$, the sets $\mathcal{F}(j)$ ensure that, for every feasible $(x,\mathbf{u})$, and for every $\check{x}$ such that $\check{x}-x \in \mathcal{F}(0)$, the condition $$\phi\big(j;\check{x},\mathbf{u},\mathbf{0}\big) \allowbreak \in \phi\big(j;x,\mathbf{u},\mathbf{0}\big) \oplus \mathcal{F}(j)$$ is verified.
	\end{itemize}
\end{assumption}
\begin{assumption}[Bounds on the effect of the perturbations on the state trajectory]
	\label{ass:cmpc_R_sequence}
	The set sequence $\{ \mathcal{R}(j) \}_{j \geq 0}$ is such that, defining $\mathcal{R}(0) \coloneqq \{ 0 \}$, it verifies the following conditions:
	\begin{enumerate}
		\item for every feasible $(x,\mathbf{u})$, the sets $\mathcal{R}(j)$, for $j > 0$, ensure $\phi\left(j;x,\mathbf{u},\mathbf{w}\right) \in \phi\left(j;x,\mathbf{u},\mathbf{0}\right) \oplus \mathcal{R}(j) \; \forall \mathbf{w} \in \mathcal{W}^j$;
		\item the set $\mathcal{X} \ominus \mathcal{R}(N_p)$ is non-empty and contains the origin;
		\item $\mathcal{F}(j) \oplus \mathcal{R}(j) \subseteq \mathcal{R}(j+1)$, $\forall j \geq 0$.
	\end{enumerate}
\end{assumption}
The sequence $\left\{ \mathcal{F}(j) \right\}_{j \geq 0}$ bounds the gap between the perturbation-free state trajectories obtained with the same control sequence starting from close enough states $x$ and $\check{x}$, while the sequence $\left\{ \mathcal{R}(j) \right\}_{j \geq 0}$ bounds the gap between the perturbed and the perturbation-free trajectories obtained starting from a common state $x$ and applying the same control sequence. 

Another ingredient that is needed to ensure the stability of the closed-loop system under our MPC is the existence of a robust controlled invariant set (RCIS) that verifies the conditions listed in the following assumption.
\begin{assumption}[Existence of an RCIS]
	\label{ass:cmpc_rcis}
	There exists a set $\Omega \subseteq \mathcal{X}$ that verifies the following conditions simultaneously:
	\begin{enumerate}
		\item for all $x \in \Omega$, $\exists u \in \mathcal{U}$ such that $f(x,u,w) \in \Omega$, for any $w \in \mathcal{W}$;
		\item $\Omega \subseteq \mathcal{X} \cap \mathcal{Q} \left( \mathcal{X} \ominus \mathcal{R}(1) \right)$ for the nominal system $\hat{x}(k+1) = f(\hat{x}(k),u(k),0)$, where $\mathcal{Q} \left( \mathcal{X} \ominus \mathcal{R}(1) \right)$ denotes the one-step controllable set to $\mathcal{X} \ominus \mathcal{R}(1)$;
		\item $\Omega \ominus \mathcal{R}(1)$ is non-empty and contains the origin.
	\end{enumerate}
\end{assumption}
\begin{remark}
	Standard MPC schemes require terminal ingredients that meet several conditions simultaneously. In detail, it is necessary to find a controlled invariant set (or an RCIS in the robust case) generated by a certain control law $\kappa_f(\cdot)$, which must also be stabilizing and cause the terminal cost function $V_f(\cdot)$ to be a CLF for the closed-loop system. 
	
	Assumption \ref{ass:cmpc_rcis} still requires to compute an RCIS, but this does not have to be related to a particular (stabilizing) control law. Moreover, it is not required to design a CLF.
\end{remark}

\subsection{Controller formulation}
\label{subsec:cmpc_controller_formulation}
Now that all the conditions that the system under control has to verify to enable the implementation of our controller have been explained, we are ready to introduce our robust contraction-based MPC, explain its design parameters and prove its stability and feasibility properties.

The control objective is assumed to be expressed by means of a stage cost function $\ell(x,u)$. Considering a generic prediction horizon $q(k)$ starting from a time instant $k$, the total cost related to the control objective is given by 
\begin{equation}
	J \big( x(k),\mathbf{u}(k),q(k) \big) \coloneqq \sum_{j=0}^{q(k)-1} \ell(\hat{x}(j \vert k),u(j \vert k)),
\end{equation}
where $\mathbf{u}(k) \coloneqq \{ u(0 \vert k),\hdots,u(q(k)-1 \vert k) \}$ and $\hat{x}(j \vert k)$ denotes the nominal $j$-step-ahead state prediction computed starting from the initial condition $x(k)$, namely $\hat{x}(j \vert k) \coloneqq \phi (j;x(k),\mathbf{u}(k),\mathbf{0})$. The stage cost function is required to fulfill the conditions listed in the next assumption.
\begin{assumption}[Stage cost function properties]
	\label{ass:cmpc_stage_cost_L}
	There exist a constant $\bar{\ell} > 0$ and a $\mathcal{K}$ function $\alpha_\ell(\cdot)$ such that
	\begin{equation}
		\label{eq:cmpc_stage_cost_upper_bound}
		\forall (x,u) \in \mathcal{X} \times \mathcal{U}, \quad 0 \leq \ell(x,u) \leq \bar{\ell},
	\end{equation}
	and $\ell(x,u) \geq \alpha_\ell(\Vert x \Vert), \; \forall u \in \mathcal{U}$. 
	
	Moreover, the stage cost function $\ell(x,u)$ is uniformly continuous with respect to all its arguments; therefore, there exist $\mathcal{K}$ functions $\sigma_{\ell,x}(\cdot)$ and $\sigma_{\ell,u}(\cdot)$ such that, $\forall x,\check{x} \in \mathcal{X}, \; u,\check{u} \in \mathcal{U}$:
	\begin{equation}
		\label{eq:cmpc_stage_cost_uniform_continuity}
		\left\vert \ell(x,u) - \ell(\check{x},\check{u}) \right\vert \leq \sigma_{\ell,x} \left( \left\Vert x-\check{x} \right\Vert \right) + \sigma_{\ell,u} \left( \left\Vert u-\check{u} \right\Vert \right).
	\end{equation}
\end{assumption}
The optimization problem $P_{N_p}\left( x(k),\theta(k) \right)$ that has to be solved at each MPC step is
\begin{subequations}
	\label{eq:cmpc_op_all_in_one}
	\begin{align}
		\min_{(\mathbf{u}(k),q(k))} &V_{N_p}\left(x(k),\theta(k),\mathbf{u}(k),q(k)\right) \coloneqq \theta(k) \cdot J\left( x(k),\mathbf{u}(k),q(k) \right) \nonumber \\
		&+ \xi \cdot \underline{\Gamma}\left( x(k),\mathbf{u}(k),q(k) \right) \nonumber \\
		\mathrm{s.t.}: \nonumber \\
		&\hat{x}(0 \vert k) = x(k), \\
		&\hat{x}(j \vert k) = f\left(\hat{x}(j-1 \vert k),u(j-1 \vert k),0\right), \; \forall j \in [1,q(k)], \\
		&\hat{x}(j \vert k) \in \mathcal{X} \ominus \mathcal{R}(j), \; \forall j \in [0,q(k)], \\
		&(\mathbf{u}(k),q(k)) \in \mathcal{U}^{q(k)} \times \left\{ 1,\hdots,N_p \right\}
	\end{align}
\end{subequations}
where:
\begin{itemize}
	\item $\mathbf{u}(k) \coloneqq \{ u(0 \vert k),\hdots,u(q(k)-1 \vert k) \}$;
	\item $q(k)$ is the prediction horizon of the controller, here considered
	as a decision variable that can take values in $[1,N_p]$;
	\item $\theta(k)$ is an internal state of the MPC controller and is
	needed to guarantee the stability properties of the
	closed-loop system and the dynamics of which is defined later.
\end{itemize}
If multiple solutions providing the same cost exist, we suppose to keep the one having the smallest $q$.
The optimal solutions of $P_{N_p}(x(k),\theta(k))$ are denoted $\mathbf{u}^*(x(k),\theta(k))$ and $q^*(x(k),\theta(k))$ (or $\mathbf{u}^*(k)$ and $q^*(k)$ for brevity) and the corresponding values $\underline{\Gamma}$, $J$, $V_{N_p}$ and $j_\mathrm{opt}$ are:
\begin{equation*}
	\begin{split}
		\underline{\Gamma}^*(x(k),\theta(k)) \coloneqq  \underline{\Gamma}(x(k),\mathbf{u}^*(k),q^*(k)),
	\end{split}
\end{equation*}
\begin{equation*}
	\begin{split}
		J^*(x(k),\theta(k)) \coloneqq  J(x(k),\mathbf{u}^*(k),q^*(k)),
	\end{split}
\end{equation*}
\begin{equation*}
	\begin{split}
		V_{N_p}^*(x(k),\theta(k)) \coloneqq  V_{N_p}(x(k),\theta(k),\mathbf{u}^*(k),q^*(k)),
	\end{split}
\end{equation*}
\begin{equation*}
	\begin{split}
		j^*(x(k),\theta(k)) \coloneqq  j_\mathrm{opt}(x(k),\mathbf{u}^*(k),q^*(k)).
	\end{split}
\end{equation*}
We also define $\hat{\mathbf{x}}^*(k) \coloneqq \{ \hat{x}^*(0 \vert k), \hdots, \hat{x}^*(q^*(k) \vert k) \}$ as the nominal state sequence provided by the optimal control sequence $\mathbf{u}^*(x(k),\theta(k))$, i.e. $\hat{x}^*(j \vert k) = \phi(j;x(k), \allowbreak \mathbf{u}^*(k),\mathbf{0})$.
The control law of the proposed MPC is the following: $\kappa_{N_p}(x(k), \theta(k)) \coloneqq  u^*(0 \vert k)$, where $u^*(0 \vert k)$ is the first element of the optimal control sequence $\mathbf{u}^*(k)$ obtained by solving $P_{N_p}(x(k),\theta(k))$.

The initial condition of the controller state $\theta$ is set to $\theta(0) =  \max \{ \epsilon, \nu \Gamma(x(0)) \}$, where $\epsilon > 0$ and $\nu \in (0,1)$ are design parameters, and is updated at every time instant $k$ by using the following policy:
\begin{equation}
	\label{eq:cmpc_h_update_rule}
	\begin{split}
		\theta(k) &= f_\theta \left( x(k),\theta(k-1) \right)
		\coloneqq \begin{cases}
			\theta(k-1), \quad \Gamma \left( x(k) \right) > \theta(k-1)\\
			\max \{ \epsilon, \nu \Gamma \left(x(k)\right) \}, \; \mathrm{otherwise}
		\end{cases}
	\end{split}
\end{equation}
Next, we want to prove that the equilibrium $x = 0$ is input-to-state practically stable (ISpS) for the closed-loop system under our robust contraction-based MPC. In order to do so, we demonstrate that $V_{N_p}^*(x(k),\theta(k))$ is an ISpS-Lyapunov function. The first step in this direction is proving that $V_{N_p}^*(x(k),\theta(k))$ is upper-bounded by a positive definite function of $x(k)$, as proved by Lemma \ref{lem:cmpc_V_{N_p}_star_ineq}, which is based on the result of Lemma \ref{lem:cmpc_V_{N_p}_star_limited}.

\begin{lemma}[Generic upper bound on the cost function value]
	\label{lem:cmpc_V_{N_p}_star_limited}
	If Assumptions \ref{ass:cmpc_continuity}-\ref{ass:cmpc_stage_cost_L} are satisfied, then for any feasible initial conditions $x(k) \in \mathcal{X}$, we have that
	\begin{equation}
		\label{eq:cmpc_V_{N_p}_star_limit}
		V_{N_p}^* \left( x(k),\theta(k) \right) \leq \theta(k)N_p\bar{\ell} + \xi \gamma \Gamma \left( x(k) \right),
	\end{equation}
	and the minimum value of the map $\Gamma(\cdot)$ is obtained at the end of the trajectory, that is
	\begin{equation}
		\label{eq:cmpc_i_star_equals_q_star}
		j^* \left( x(k),\theta(k) \right) = q^* \left( x(k),\theta(k) \right),
	\end{equation}
	where $j^*$ is the horizon providing the maximum contraction.
\end{lemma}
\begin{proof}
	The proof is provided in the Appendix \ref{subsec:proofs}.
\end{proof}

\begin{lemma}[Existence of an upper-bounding function for the cost function]
	\label{lem:cmpc_V_{N_p}_star_ineq}
	Under Assumptions \ref{ass:cmpc_continuity}-\ref{ass:cmpc_stage_cost_L}, given a feasible initial condition $x(k) \in \mathcal{X}$, if the following conditions hold simultaneously:
	\begin{enumerate}
		\item $\theta(k) = f_\theta \left( x(k),\theta(k-1) \right)$,
		\item $\xi \geq 2N_p\bar{\ell} / (1-\gamma)$,
	\end{enumerate}
	then the optimal solution to $P_{N_p}(x(k),\theta(k))$ satisfies the inequality
	\begin{equation}
		\label{eq:cmpc_V_{N_p}_star_upper_bound}
		V_{N_p}^* \left( x(k),\theta(k) \right) \leq \left[ \frac{1+\gamma}{2} \right] \xi \Gamma \left( x(k) \right) + \epsilon N_p \bar{\ell}.
	\end{equation}
\end{lemma}
\begin{proof}
	The proof is provided in the Appendix \ref{subsec:proofs}.
\end{proof}
Next, we prove the nominal descent property of $V_{N_p}^* \left( x(k),\theta(k) \right)$ and the boundedness of its increase due to the uncertainty $w(k)$, discerning the cases where $q^*(x(k), \allowbreak \theta(k)) > 1$ (Lemma \ref{lem:cmpc_q_greater_1}) and $q^*(x(k),\theta(k)) = 1$ (Lemma \ref{lem:cmpc_q_equal_1}).

\begin{lemma}[Recursive feasibility and cost descent for $q^* > 1$]
	\label{lem:cmpc_q_greater_1}
	Given a feasible initial condition $ x(k) \in \mathcal{X}$, under Assumptions \ref{ass:cmpc_continuity}-\ref{ass:cmpc_stage_cost_L}, if the following conditions hold simultaneously:
	\begin{enumerate}
		\item $q^* \left( x(k),\theta(k) \right) > 1$,
		\item $\theta(k)=f_\theta \big( x(k),\theta(k-1) \big)$,
	\end{enumerate}
	then $P_{N_p} \left( x(k+1), \theta(k+1) \right)$ is feasible, and there exists a $\mathcal{K}$ function $\lambda_1(\cdot)$ such that
	\begin{equation}
		\label{eq:cmpc_1_ISS_Lyapunov}
		\begin{split}
			&V_{N_p} \left( x(k+1),\theta(k+1),\mathbf{u}^+(k+1),q^+(k+1) \right) - V_{N_p}^* \left( x(k),\theta(k) \right)\\ &\leq -\theta(k) \alpha_\ell \left( \Vert x(k) \Vert \right) + \left( \theta(k) + \xi \right) \lambda_1 \left( \Vert w(k) \Vert \right),
		\end{split}
	\end{equation}
	where $q^+(k+1) \coloneqq q^*(k) - 1$ and $\mathbf{u}^+(k+1) \coloneqq \big\{ u^*(1 \vert k),\hdots,u^*(q^*(k)-1 \vert k) \big\}$.
\end{lemma}
\begin{proof}
	The proof is provided in the Appendix \ref{subsec:proofs}.
\end{proof}
Before introducing the next lemma, we define $\omega > 0$ as a constant such that $\{ x: \; \Gamma(x) \leq \omega \} \subseteq \Omega \ominus \mathcal{R}(1)$.
\begin{lemma}[Recursive feasibility and cost descent for $q^* = 1$]
	\label{lem:cmpc_q_equal_1}
	Under Assumptions \ref{ass:cmpc_continuity}-\ref{ass:cmpc_stage_cost_L}, given a feasible initial condition $x(k) \in \mathcal{X}$, if the following conditions hold simultaneously:
	\begin{enumerate}
		\item $q^* \left( x(k),\theta(k) \right) = 1$,
		\item $\theta(k) = f_\theta \big( x(k),\theta(k-1) \big)$,
		\item $\xi \geq 2N_p\bar{\ell}/(1-\gamma)$,
		\item $\gamma \leq \frac{\omega}{\Gamma_\mathrm{max}}$, where $\Gamma_\mathrm{max} \coloneqq \max_{x \in \mathcal{X}} \Gamma(x)$,
	\end{enumerate}
	then $P_{N_p} \left( x(k+1),\theta(k+1) \right)$ is feasible, and there exists a $\mathcal{K}$ function $\lambda_2(\cdot)$ such that the following inequality holds true:
	\begin{equation}
		\label{eq:cmpc_2_ISS_Lyapunov}
		\begin{split}
			&V_{N_p}^* \left( x(k+1),\theta(k+1) \right) - V_{N_p}^* \left( x(k),\theta(k) \right) \leq -\theta(k) \alpha_\ell \left( \Vert x(k) \Vert \right)\\ &+ \lambda_2 \left( \Vert w(k) \Vert \right) + \epsilon N_p \bar{\ell}.
		\end{split}
	\end{equation}
\end{lemma}
\begin{proof}
	The proof is provided in the Appendix \ref{subsec:proofs}.
\end{proof}

We are now ready to present the main theorem of the work, which proves input-to-state practical stability of the closed loop and recursive feasibility of our robust contraction-based MPC.

\begin{theorem}[Stability of the robust contraction-based MPC for the regulation of nonlinear systems]
	\label{thm:cmpc}
	Under Assumptions \ref{ass:cmpc_continuity}-\ref{ass:cmpc_stage_cost_L}, if the following conditions hold simultaneously:
	\begin{enumerate}
		\item $\xi \geq 2N_p\bar{\ell}/(1-\gamma)$,
		\item $\theta(k) = f_\theta \left( x(k),\theta(k-1) \right)$,
		\item $\gamma \leq \frac{\omega}{\Gamma_\mathrm{max}}$,
	\end{enumerate}
	then, given a feasible initial condition $x \in \mathcal{X}$, $P_{N_p} \left( x,\theta \right)$ is recursively feasible and the equilibrium $x = 0$ is ISpS for the closed-loop system.
\end{theorem}
\begin{proof}
	The proof is provided in the Appendix \ref{subsec:proofs}.
\end{proof}
\begin{remark}
	It is important to note that the input-to-state practical stability property ensured by Theorem \ref{thm:cmpc} is very close to input-to-state stability. Indeed, $V_{N_p}^*(\cdot,\cdot)$ is an ISpS-Lyapunov function because in its upper bound function and in its descent inequality some positive constant terms appear, but these constant terms all depend on $\epsilon$, which can be set to very small values. Therefore, setting $\epsilon$ to values that are close to zero, $V_{N_p}^*(\cdot,\cdot)$ practically behaves as an ISS-Lyapunov function.
\end{remark}

\subsection{A robust contraction-based MPC formulation without integer decision variables}
\label{sec:cmpc_2op_cmpc}
The MPC formulation introduced in \eqref{eq:cmpc_op_all_in_one} involves the integer variable $q$, which could make the resolution of the problem computationally demanding. In this section, an alternative two-stage algorithm without integer decision variables is introduced.

The two steps of the newly proposed algorithm are the following:
\begin{enumerate}
	\item First, the following fixed-horizon problem is solved:
	\begin{equation}
		\label{eq:cmpc_op1}
		\begin{split}
			\min_{\mathbf{u}(k)} \; &\underline{\Gamma}(x(k),\mathbf{u}(k),N_p)\\
			\mathrm{s.t.} \; &\hat{x}(0 \vert k) = x(k), \\
			&\hat{x}(j \vert k) = f\left(\hat{x}(j   -   1 \vert k),u(j-1 \vert k),0\right),  \forall j \in [1,N_p], \\
			&\hat{x}(j \vert k) \in \mathcal{X} \ominus \mathcal{R}(j), \; \forall j \in [0,N_p], \\
			&\mathbf{u}(k) \in \mathcal{U}^{N_p} 
		\end{split}
	\end{equation}
	and the instant $j_{N_p} \in [1,N_p]$ when the maximum contraction occurs is memorized.
	\item A second optimization problem with prediction horizon equal to $j_{N_p}$ is solved to get the desired optimal control sequence:
	\begin{equation}
		\label{eq:cmpc_op2}
		\begin{split}
			\min_{\mathbf{u}(k)} \; &V_{N_p}\left(x(k),\theta(k),\mathbf{u}(k),j_{N_p}\right)\\
			\mathrm{s.t.} \; &\hat{x}(0 \vert k) = x(k), \\
			&\hat{x}(j \vert k)   =   f\left(\hat{x}(j   -  1 \vert k),u(j-1 \vert k),0\right), \forall j \in [1,j_{N_p}], \\
			&\hat{x}(j \vert k) \in \mathcal{X} \ominus \mathcal{R}(j), \; \forall j \in [0,j_{N_p}], \\
			&\mathbf{u}(k) \in \mathcal{U}^{j_{N_p}}. 
		\end{split}
	\end{equation}
\end{enumerate}
This formulation inherits all the stability and robustness properties of the one
introduced before, as proved by the next theorem.
\begin{theorem}[Stability of the two-stage formulation of the robust contraction-based MPC for the regulation of nonlinear systems]
	\label{thm:cmpc_2op}
	If the following conditions hold:
	\begin{enumerate}
		\item Assumptions \ref{ass:cmpc_continuity}-\ref{ass:cmpc_stage_cost_L} are satisfied;
		\item the penalty $\xi$ involved in the second optimization problem satisfies $\xi \geq 2N_p\bar{\ell}/(1-\gamma)$;
		\item the rule $\theta(k) = f_\theta (x(k),\theta(k-1))$ is used to update the value of $\theta$,
		\item $\gamma \leq \frac{\omega}{\Gamma_\mathrm{max}}$,
	\end{enumerate}
	then, given a feasible initial condition $x \in \mathcal{X}$, the 2-stage version of $P_{N_p} \left( x,\theta \right)$ is recursively feasible and the equilibrium $x = 0$ is ISpS for the closed-loop system.
\end{theorem}
\begin{proof}
	The proof is provided in the Appendix \ref{subsec:proofs}.
\end{proof}
\begin{remark}
	In order to limit the increment in the computational
	cost provided by this formulation, the first $j_{N_p}$ elements of the solution of problem \eqref{eq:cmpc_op1} could be used as feasible initial solution for problem \eqref{eq:cmpc_op2}.

	Moreover, it is important to note that also with this formulation choosing a very small $\epsilon$ renders $V_{N_p}^*(\cdot,\cdot)$ (almost) an ISS-Lyapunov function.
\end{remark}

\section{Case studies}
\label{sec:case_studies}
In order to validate our robust contraction-based MPC, we employ it to regulate two different systems and check its effectiveness. In particular, the aims of the following case studies are two: (i) prove how our MPC can control systems for which common methodologies for designing the terminal ingredients are ineffective, and (ii) prove how our MPC can also be implemented starting from terminal ingredients that are designed following standard state-of-the-art techniques.

In detail, we first control a perturbed nonholonomic system, and then control a perturbed four-tank system. In both cases, we rely on the 2-stage version of our robust contraction-based MPC.

\subsection{Control of a perturbed nonholonomic system}
\label{subsec:nonholonomic_system}
We here consider a modified version of the illustrative example provided in \cite{alamir2017contraction}. Precisely, the aim is to control the following perturbed discrete-time nonholonomic system to the origin starting from the initial condition $x(0) = (-4,10,4)$:
\begin{subequations}
	\label{eq:nonholonomic_system}
	\begin{equation}
		x_1(k+1) = x_1(k) + \left( 1 + w(k) \right)u_1(k)
	\end{equation}
	\begin{equation}
		x_2(k+1) = x_2(k) + u_2(k)
	\end{equation}
	\begin{equation}
		x_3(k+1) = x_3(k) + x_1(k)u_2(k).
	\end{equation}
\end{subequations}
In detail, $x \coloneqq (x_1,x_2,x_3) \in \mathbb{R}^3$ is the state of the system, $u \coloneqq (u_1,u_2) \in \mathbb{R}^2$ is the input, and $w$ is a perturbation that is assumed to be bounded. In particular, $w \in \mathcal{W} \coloneqq [-0.025,0.025]$. In the following, we will often refer to this using the simplified notation $x(k+1) = f \left( x(k),u(k),w(k) \right)$.

The state and the input of the system are subject to the following constraints: 
\begin{equation*}
	x(k) \in \mathcal{X} \coloneqq \left\{ x \in \mathbb{R}^3: \; \vert x_1 \vert \leq 4, \vert x_2 \vert \leq 10,\; \vert x_3 \vert \leq 10 \right\}, \; \forall k \geq 0,
\end{equation*}
\begin{equation*}
	u(k) \in \mathcal{U} \coloneqq \left\{ u \in \mathbb{R}^2: \; \vert u_1 \vert \leq 8, \vert u_2 \vert \leq 0.5 \right\}, \; \forall k \geq 0.
\end{equation*}
Before proceeding with the design of our robust contraction-based MPC, let us analyze why the design of a standard MPC equipped with a terminal cost function and a terminal inequality constraint would be challenging. As the classical methods to design such terminal ingredients require to linearize the nominal system, i.e. setting $w = 0$, at the reference equilibrium, we proceed by linearizing the system at the origin, obtaining the following state and input matrices:
\begin{equation*}
	A = \begin{bmatrix}
		1 & 0 & 0 \\ 0 & 1 & 0 \\ 0 & 0 & 1
	\end{bmatrix}, \; B = \begin{bmatrix}
	1 & 0 \\ 0 & 1 \\ 0 & 0
	\end{bmatrix}.
\end{equation*}
The previous matrices clearly show that the last state of the linearized system is uncontrollable and such that it remains constantly equal to its initial value. Therefore, it is practically impossible to resort to standard methods for designing a locally stabilizing controller and a terminal cost function that locally behaves as a CLF, e.g. by solving an LMI based on the matrices $A$ and $B$ to obtain a linear controller $u = Kx$ and a quadratic terminal cost weight matrix $P$ that satisfy all the needed conditions. 

In cases like this, one could consider to act differently and opt for MPC formulations with terminal equality constraints, which make it possible to ensure the stability of the closed loop also without knowing a locally stabilizing control law and a local CLF. Nonetheless, implementing a tube-based MPC with terminal equality constraint is quite unusual. In fact, ensuring the feasibility of a terminal equality constraint MPC often requires to have longer prediction horizons, but at the same time tube-based MPCs need short prediction horizons, as the sequences of tightened constraints that they employ to ensure robustness could lead to an empty feasible set for longer prediction horizons.

For these reasons, we here employ our robust contraction-based MPC to show its effectiveness. In the following, we discuss (i) the computation of the sequences $\{\mathcal{F}(j)\}_{j \geq 0}$ and $\{\mathcal{R}(j)\}_{j \geq 0}$, (ii) the computation of the RCIS $\Omega$, (iii) the consequent choice of the contractive function and the maximum prediction horizon, and (iv) the choice of the MPC settings. Finally, we show the results of our simulations.

\subsubsection{Computation of the sequences $\{\mathcal{F}(j)\}_{j \geq 0}$ and $\{\mathcal{R}(j)\}_{j \geq 0}$}
\label{subsubsec:nhs_F_R}
Before delving into the technical details behind the computation of the sequences $\{\mathcal{F}(j)\}_{j \geq 0}$ and $\{\mathcal{R}(j)\}_{j \geq 0}$, we first note that system \eqref{eq:nonholonomic_system} is Lipschitz in the compact set $\mathcal{X} \times \mathcal{U} \times \mathcal{W}$. However, as reported in \cite{Manzano2019_Choki}, the Lipschitz continuity in a compact set implies also component-wise Lipschitz continuity in the same set, namely there exist non-negative real constants $L_{x,ia}$, $L_{u,ib}$ and $L_{w,i}$, where $i,a \in \{1,2,3\}$, $b \in \{1,2\}$, such that
\begin{equation}
	\label{eq:comp_Lipschitz_nhs}
	\begin{split}
		\vert f_i(x,u,w) - f_i(\check{x},\check{u},\check{w}) \vert &\leq \sum_{a = 1}^3 L_{x,ia} \vert x_a - \check{x}_a \vert + \sum_{b = 1}^2 L_{u,ib} \vert u_b - \check{u}_b \vert\\ &+ L_{w,i} \vert w - \check{w} \vert.
	\end{split}
\end{equation}
In particular, the values of the constants $L_{x,ia}$, $L_{u,ib}$ and $L_{w,i}$ are respectively reported in Tables \ref{tab:cmpc_Lx}, \ref{tab:cmpc_Lu} and \ref{tab:cmpc_Lw}.
\begin{table}[h]
	\centering
	\arrayrulecolor{black}
	\begin{tabular}{|c|c|c|c|c|} 
		\hline
		\rowcolor{myblue!10} \multicolumn{2}{|c|}{{\cellcolor{myblue!10}}} & \multicolumn{3}{c|}{$a$} \\ 
		\hhline{|>{\arrayrulecolor{myblue!10}}-->{\arrayrulecolor{black}}---|}
		\rowcolor{myblue!10} \multicolumn{2}{|c|}{\multirow{-2}{*}{{\cellcolor{myblue!10}}$L_{x,ia}$}} & \textit{1} & \textit{2} & \textit{3} \\ 
		\hline
		{\cellcolor{myblue!10}} & {\cellcolor{myblue!10}}\textit{1} & 1 & 0 & 0 \\ 
		\hhline{|>{\arrayrulecolor{myblue!10}}->{\arrayrulecolor{black}}----|}
		{\cellcolor{myblue!10}} & {\cellcolor{myblue!10}}\textit{2} & 0 & 1 & 0 \\ 
		\hhline{|>{\arrayrulecolor{myblue!10}}->{\arrayrulecolor{black}}----|}
		\multirow{-3}{*}{{\cellcolor{myblue!10}}$i$} & {\cellcolor{myblue!10}}\textit{3} & 0.5 & 0 & 1 \\
		\hline
	\end{tabular}
	\caption{Values of the constants $L_{x,ia}$ of the nonholonomic system.}
	\label{tab:cmpc_Lx}
\end{table}

\begin{table}[h]
	\centering
	\arrayrulecolor{black}
	\begin{tabular}{|c|c|c|c|} 
		\hline
		\rowcolor{myblue!10} \multicolumn{2}{|c|}{{\cellcolor{myblue!10}}} & \multicolumn{2}{c|}{$b$} \\ 
		\hhline{|>{\arrayrulecolor{myblue!10}}-->{\arrayrulecolor{black}}--|}
		\rowcolor{myblue!10} \multicolumn{2}{|c|}{\multirow{-2}{*}{{\cellcolor{myblue!10}}$L_{u,ib}$}} & \textit{1} & \textit{2} \\ 
		\hline
		{\cellcolor{myblue!10}} & {\cellcolor{myblue!10}}\textit{1} & 1.025 & 0 \\ 
		\hhline{|>{\arrayrulecolor{myblue!10}}->{\arrayrulecolor{black}}---|}
		{\cellcolor{myblue!10}} & {\cellcolor{myblue!10}}\textit{2} & 0 & 1 \\ 
		\hhline{|>{\arrayrulecolor{myblue!10}}->{\arrayrulecolor{black}}---|}
		\multirow{-3}{*}{{\cellcolor{myblue!10}}$i$} & {\cellcolor{myblue!10}}\textit{3} & 0 & 4 \\
		\hline
	\end{tabular}
	\caption{Values of the constants $L_{u,ib}$ of the nonholonomic system.}
	\label{tab:cmpc_Lu}
\end{table}
\begin{table}[h]
	\centering
	\begin{tabular}{|c|c|c|}
		\hline
		\rowcolor{myblue!10}
		$L_{w,1}$ & $L_{w,2}$ & $L_{w,3}$ \\ \hline
		8 & 0 & 0 \\ \hline
	\end{tabular}
	\caption{Values of the constants $L_{w,i}$ of the nonholonomic system..}
	\label{tab:cmpc_Lw}
\end{table}

Provided that component-wise Lipschitz continuity is a stronger condition than component-wise uniform continuity, our robust contraction-based MPC can be applied to system \eqref{eq:nonholonomic_system}. Moreover, the component-wise Lipschitz continuity of the system makes it possible to exploit Algorithm \ref{subsec:F_R_Lipschitz} for the computation of the sequences $\{\mathcal{F}(j)\}_{j \geq 0}$ and $\{\mathcal{R}(j)\}_{j \geq 0}$. As the computed sets are box-shaped, i.e.
\begin{equation*}
	\mathcal{F}(j) = \{ x \in \mathbb{R}^3: \; \vert x_i \vert \leq \bar{x}_{i,\mathcal{F}(j)}, \forall i \in \{1,2,3\} \}
\end{equation*}
and
\begin{equation*}
	\mathcal{R}(j) = \{ x \in \mathbb{R}^3: \; \vert x_i \vert \leq \bar{x}_{i,\mathcal{R}(j)}, \forall i \in \{1,2,3\} \},
\end{equation*}
we provide the computed $\bar{x}_{i,\mathcal{F}(j)}$ and $\bar{x}_{i,\mathcal{R}(j)}$ in Table \ref{tab:cmpc_F_R_values}.
\begin{table}
	\centering
	\arrayrulecolor{black}
	\resizebox{\columnwidth}{!}{%
	\begin{tabular}{|c|c|c|c|c|c|c|c|c|c|c|c|c|} 
		\hline
		\rowcolor{myblue!10} \multicolumn{2}{|c|}{$j$} & \textit{0} & \textit{1} & \textit{2} & \textit{3} & \textit{4} & \textit{5} & \textit{6} & \textit{7} & \textit{8} & \textit{9} & \textit{10} \\ 
		\hline
		{\cellcolor{myblue!10}} & {\cellcolor{myblue!10}}$\bar{x}_{1,\mathcal{F}(j)}$ & 0.2 & 0.2 & 0.2 & 0.2 & 0.2 & 0.2 & 0.2 & 0.2 & 0.2 & 0.2 & 0.2 \\ 
		\hhline{|>{\arrayrulecolor{myblue!10}}->{\arrayrulecolor{black}}------------|}
		{\cellcolor{myblue!10}} & {\cellcolor{myblue!10}}$\bar{x}_{2,\mathcal{F}(j)}$ & 0 & 0 & 0 & 0 & 0 & 0 & 0 & 0 & 0 & 0 & 0 \\ 
		\hhline{|>{\arrayrulecolor{myblue!10}}->{\arrayrulecolor{black}}------------|}
		\multirow{-3}{*}{{\cellcolor{myblue!10}}$\mathcal{F}(j)$} & {\cellcolor{myblue!10}}$\bar{x}_{3,\mathcal{F}(j)}$ & 0 & 0.1 & 0.2 & 0.3 & 0.4 & 0.5 & 0.6 & 0.7 & 0.8 & 0.9 & 1.0 \\ 
		\hline
		{\cellcolor{myblue!10}} & {\cellcolor{myblue!10}}$\bar{x}_{1,\mathcal{R}(j)}$ & 0 & 0.2 & 0.4 & 0.6 & 0.8 & 1.0 & 1.2 & 1.4 & 1.6 & 1.8 & 2.0 \\ 
		\hhline{|>{\arrayrulecolor{myblue!10}}->{\arrayrulecolor{black}}------------|}
		{\cellcolor{myblue!10}} & {\cellcolor{myblue!10}}$\bar{x}_{2,\mathcal{R}(j)}$ & 0 & 0 & 0 & 0 & 0 & 0 & 0 & 0 & 0 & 0 & 0 \\ 
		\hhline{|>{\arrayrulecolor{myblue!10}}->{\arrayrulecolor{black}}------------|}
		\multirow{-3}{*}{{\cellcolor{myblue!10}}$\mathcal{R}(j)$} & {\cellcolor{myblue!10}}$\bar{x}_{3,\mathcal{R}(j)}$ & 0 & 0 & 0.1 & 0.3 & 0.6 & 1.0 & 1.5 & 2.1 & 2.8 & 3.6 & 4.5 \\
		\hline
	\end{tabular}%
	}
	\caption{Description of the sequences $\{ \mathcal{F}(j) \}_{j \geq 0}$ and $\{ \mathcal{R}(j) \}_{j \geq 0}$ that were obtained by means of the algorithm presented in the Appendix \ref{subsec:F_R_Lipschitz}.}
	\label{tab:cmpc_F_R_values}
\end{table}
As the tightened constraint sets employed in our MPC are of the kind $\mathcal{X} \ominus \mathcal{R}(j)$, it is evident from Table \ref{tab:cmpc_F_R_values} that the state that suffers most from the propagation of uncertainty is $x_3$, and that a maximum prediction horizon $N_p > 10$ would lead to an empty feasible set.

\subsubsection{Computation of a robust controlled invariant set}
As clarified in Section \ref{sec:robust_cmpc}, the knowledge of an RCIS $\Omega$ is fundamental to guarantee the recursive feasibility of our robust contraction-based MPC. In detail, it is needed to compute the maximum value that the contraction factor $\gamma$ can take to guarantee the recursive feasibility of our MPC when the chosen prediction horizon is $q = 1$. 

In the case of system \eqref{eq:nonholonomic_system}, the computation of an RCIS is straightforward. Note, indeed, that for any $x \in \mathcal{X}$ the input $u = (0,0)$ is feasible and such that $f(x,u,w) = x$, $\forall w \in \mathcal{W}$. Therefore, as $\Omega$ is only required to be robustly invariant, without the need of finding a locally stibilizing control law, the last finding is sufficient to conclude that the entire state constraint set $\mathcal{X}$ is an RCIS, i.e. $\Omega \equiv \mathcal{X}$.

\subsubsection{Choice of the contractive function and the maximum prediction horizon}
\label{subsubsec:contractive_function}
The computed sequence $\{\mathcal{R}(j)\}_{j \geq 0}$ puts an upper bound on the maximum prediction horizon that we can set. In particular, we must have that $N_p \leq 10$. Therefore, we must find a contractive positive definite function $\Gamma(x)$ that contracts sufficiently fast to ensure that the set $\Omega \ominus \mathcal{R}(1)$ is reached by the nominal system within 10 steps starting from any initial condition $x \in \mathcal{X}$.

While from a theoretical point of view the choice of the best contractive function candidate is still a subject to be investigated, what we have done in the context of the current example is to choose a positive definite function that could maximize the volume of the sublevel set $\{ \Gamma(x) \leq \omega \}$, which must lie inside the set $\Omega \ominus \mathcal{R}(1) \equiv \mathcal{X} \ominus \mathcal{R}(1)$. The intuition behind this idea is that by maximizing the volume of such set we could potentially obtain large values of $\omega$, and consequently get a larger upper bound for the contraction factor $\gamma$ within the horizon $[1,10]$.
\begin{remark}
	The last consideration does not need to be true in every situation. In fact, given that the upper bound of $\gamma$ is computed as $\omega/\Gamma_\mathrm{max}$, where $\Gamma_\mathrm{max}$ is the maximum value that is taken by $\Gamma(x)$ in $\mathcal{X}$, the shape and the volume of the sets $\mathcal{X}$ and $\Omega$ have to be taken into account to choose a suitable contractive function $\Gamma(x)$. In addition, as the contractivity of a candidate function depends also on the dynamics of the system, it might also happen that the candidate function $\Gamma(x)$ maximizing the volume of $\{ \Gamma(x) \leq \omega \}$ actually results in a smaller upper bound for $\gamma$, due to the dynamics of the system not favouring the descent of $\Gamma(x)$.
	
	In any case, should a contractive function $\Gamma(x)$ lead to an unsatisfying contraction factor $\gamma$ within the desired horizon $[1,N_p]$, it is always possible to use a state constraint set $\hat{\mathcal{X}} \subset \mathcal{X}$ that can be shaped to reduce the value of $\Gamma_\mathrm{max}$. This would reduce the space where the system state is allowed to move, but would keep the stabilizing properties of the here-proposed robust contraction-based MPC.
\end{remark}
After the last considerations, we are ready to present the employed contractive function:
\begin{equation}
	\Gamma(x) = x^\top P x, \quad P = \begin{bmatrix}
		1 & 0 & 0 \\ 0 & 0.167 & 0 \\ 0 & 0 & 0.167
	\end{bmatrix}.
\end{equation}
The employment of this contractive function results in $\omega = 14.44$ and $\Gamma_\mathrm{max} = 49.4$. Therefore, the upper bound of the contractivity factor is given by $\omega / \Gamma_\mathrm{max} = 0.2923$.

In order to estimate the real contraction factor $\gamma$, we act as follows:
\begin{enumerate}
	\item a grid of 8000 points in $\mathcal{X}$ is created;
	\item for each $\hat{N}_p \in [1,10]$ and for each sample $x \in \mathcal{X}$, the punctual contraction factor $\gamma(\hat{N}_p,x)$ is computed by solving the following optimization problem:
	\begin{equation*}
		\label{eq:op_contraction}
		\begin{split}
			\min_{\mathbf{u}} \; &\Gamma \left( \hat{x}(\hat{N}_p) \right) / \Gamma \left( \hat{x}(0) \right)\\
			\mathrm{s.t.} \; &\hat{x}(0) = x, \\
			&\hat{x}(j) = f\left(\hat{x}(j - 1),u(k + j - 1),0\right),  \forall j \in [1,\hat{N}_p], \\
			&\mathbf{u} \in \mathcal{U}^{\hat{N}_p} 
		\end{split}
	\end{equation*}
	\item for each $\hat{N}_p \in [1,10]$, the respective contraction factor $\gamma(\hat{N}_p)$ is computed as $$\gamma(\hat{N}_p) = \max_{x \in \mathcal{X}} \gamma(\hat{N}_p,x);$$
	\item the maximum prediction horizon of our MPC is chosen as $$N_p = \min_{\hat{N}_p \in [1,10]} \hat{N}_p \; \mathrm{s.t.} \gamma(\hat{N}_p) \leq \omega/\Gamma_\mathrm{max};$$
	\item the final contraction factor is set as $\gamma = \gamma(N_p)$.
\end{enumerate}
Following this algorithm, we get a maximum prediction horizon $N_p = 10$ and a corresponding contraction factor $\gamma = 0.2487$, which is lower than the previously computed upper bound.
\begin{remark}
	Note that the optimization problem that is solved for each prediction horizon and for each sample $x$ does not impose the predicted states to belong to the set $\mathcal{X}$. This is because the theory presented in Section \ref{sec:robust_cmpc} does not require this, as it ensures recursive feasibility starting from a feasible initial condition and does not try to give an analytical formulation of the feasibility set. Of course, this means that it is possible to find a contraction factor $\gamma$ verifying all the conditions listed in Section \ref{sec:robust_cmpc} and have no initial conditions that make our robust contraction-based MPC feasible.
\end{remark}

\subsubsection{Controller settings}
Now that the main ingredients of our MPC have been defined, we need to choose a stage cost function and set the parameters $\nu$, $\epsilon$, $\xi$, and the initial condition of the internal controller state $\theta(0)$.

The stage cost function $\ell(x,u)$ is chosen as a quadratic cost function $\ell(x,u) = x^\top Q x + u^\top R u$, where $Q = I_3$ and $R = 0.01I_2$, resulting in $\bar{\ell} = 216.6425$. The parameter $\xi$ is chosen as $\xi = 5.7668 \cdot 10^3$ to verify the conditions of Theorem \ref{thm:cmpc_2op}, while $\nu$ and $\epsilon$, that are the parameters used in the update rule for $\theta(k)$, are set to $\nu = 0.99$ and $\epsilon = 10^{-8}$. The initial condition of the internal state $\theta$ is set to $\theta(0) = \nu\Gamma\left( x(0) \right) = 35.0183$.

\subsubsection{Simulation results}
In order to emphasize the robustness of the proposed method under different perturbation realizations, we run 100 simulations of the closed-loop system, setting the length of each simulation to 30 steps. All the simulations start from the given initial condition and use a perturbation $w$ that is randomly sampled from $\mathcal{W}$ at each instant $k$. Our robust contraction-based MPC is implemented in its two-stage formulation.

\begin{figure}[h]
	\includegraphics[trim=3cm 7cm 3cm 7cm, clip,width=\columnwidth]{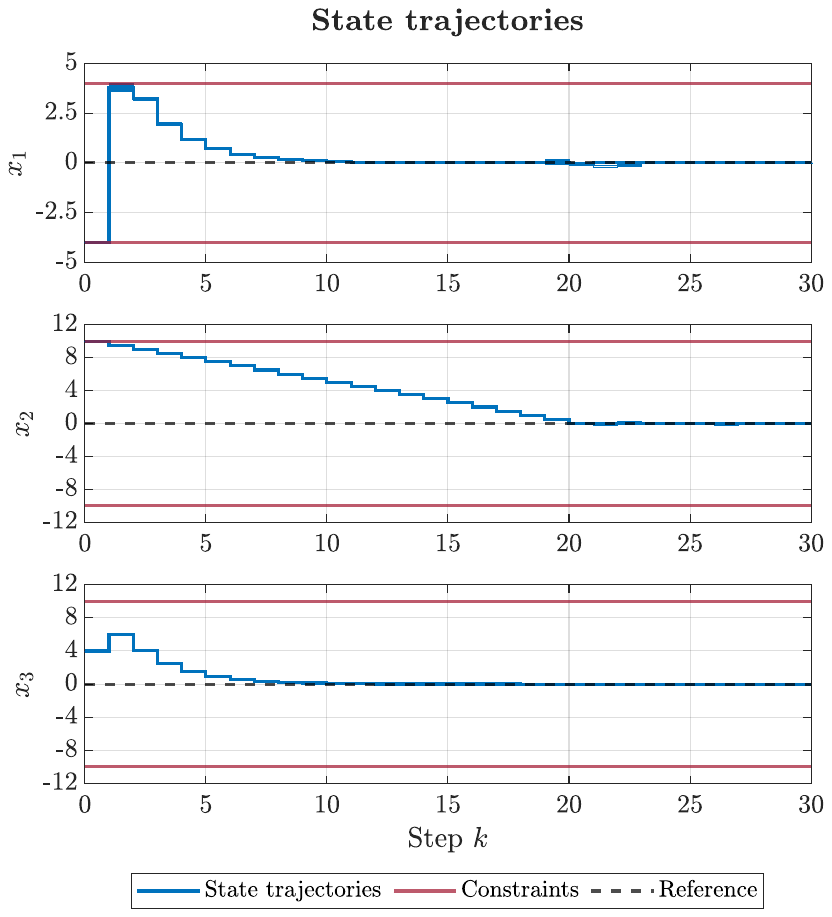}
	\centering
	\caption{Closed-loop state trajectories under our robust contraction-based MPC.} 
	\label{fig:states_nhs}
\end{figure}

\begin{figure}[h]
	\includegraphics[trim=3cm 9.75cm 3cm 9cm, clip,width=\columnwidth]{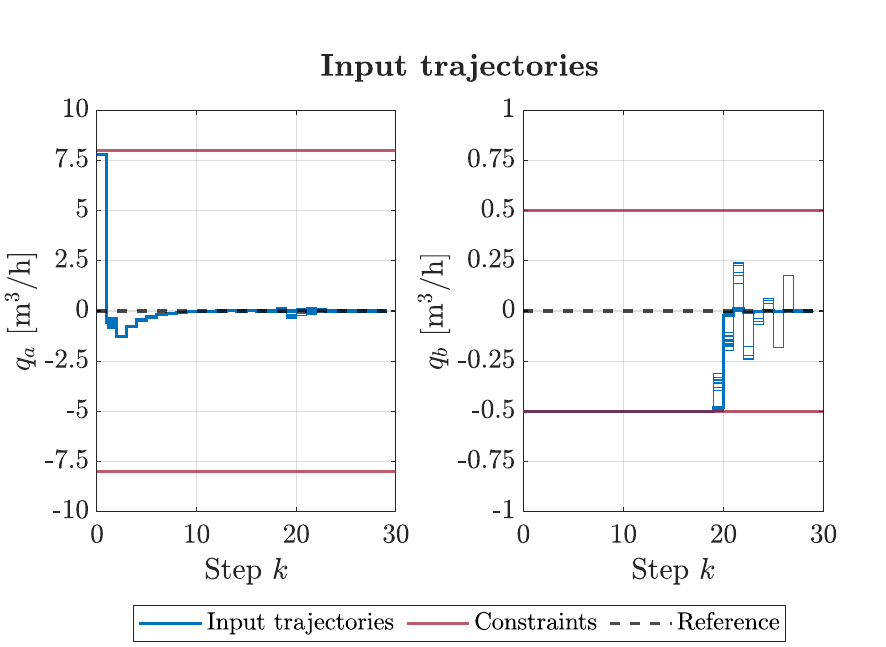}
	\centering
	\caption{Control actions provided by our robust contraction-based MPC.} 
	\label{fig:inputs_nhs}
\end{figure}

The trajectories of the closed-loop system state are shown in Figure \ref{fig:states_nhs}, which gives evidence of the stabilizing properties of our robust contraction-based MPC. Moreover, note that the state always verifies the constraints regardless of the perturbation realizations. Also the control actions provided by our MPC always verify the input constraints, as proved by Figure \ref{fig:inputs_nhs}. 

\begin{figure}[h]
	\includegraphics[trim=3cm 10cm 3cm 9cm, clip,width=\columnwidth]{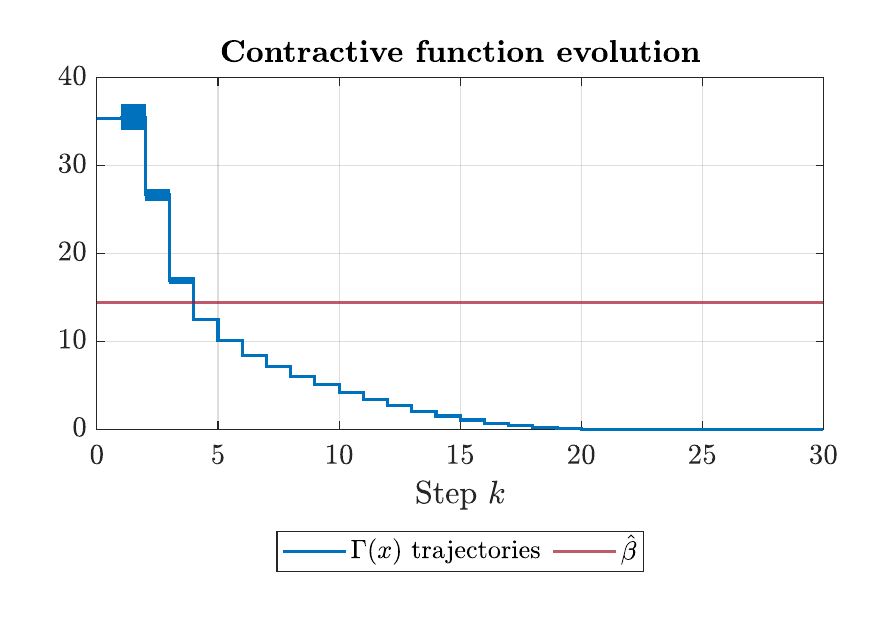}
	\centering
	\caption{Evolution of the contractive function $\Gamma(x)$ during the simulations.} 
	\label{fig:Gamma_nhs}
\end{figure}

It is also interesting to analyze the evolution of the contractive function $\Gamma(x)$ over time, which is shown in Figure \ref{fig:Gamma_nhs}. Indeed, it is possible to note how the value of such contractive function actually grows sometimes. Nevertheless, this is not inconsistent with the theory. Our robust contraction-based MPC requires the nominal system, i.e. without perturbations, to have the contractivity property. It should also be noted that the sublevel set $\{x: \; \Gamma(x) \leq \omega\}$ is reached after only 4 steps on average. This is consistent with the fact that $\Omega \equiv \mathcal{X}$ and that the chosen contractive function maximizes the volume of the sublevel set $\{x: \; \Gamma(x) \leq \omega\}$. 

\subsection{Control of a perturbed four-tank system}
\label{subsec:four_tank_system}
Section \ref{subsec:nonholonomic_system} provides an example that shows how our robust contraction-based MPC is a good alternative to terminal equality constraints in those cases in which the computation of standard terminal ingredients is difficult. In the following, we will instead show how the here-proposed MPC can be used also when standard terminal ingredients can be computed, and we will give proof of how such ingredients can actually be exploited during the design of our robust contraction-based MPC.

The considered system is a forward Euler discretization of a perturbed quadruple tank system, which is described by the following equations:
\begin{subequations}
	\label{eq:quadruple_tank}
	\begin{equation}
		\begin{split}
			h_1(k+1) &= h_1(k) - \frac{a_1T_s}{S}\sqrt{2gh_1(k)} + \frac{a_3T_s}{S}\sqrt{2gh_3(k)}\\ &+ \frac{(\gamma_1 + w_1(k))T_s}{3600S}q_1(k),
		\end{split}
	\end{equation}
	\begin{equation}
		\begin{split}
			h_2(k+1) &= h_2(k) - \frac{a_2T_s}{S}\sqrt{2gh_2(k)} + \frac{a_4T_s}{S}\sqrt{2gh_4(k)}\\ &+ \frac{(\gamma_2 + w_2(k))T_s}{3600S}q_2(k),
		\end{split}
	\end{equation}
	\begin{equation}
		\begin{split}
			h_3(k+1) &= h_3(k) - \frac{a_3T_s}{S}\sqrt{2gh_3(k)} + \frac{(1 - \gamma_2 - w_2(k))T_s}{3600S}q_2(k),
		\end{split}
	\end{equation}
	\begin{equation}
		\begin{split}
			h_4(k+1) &= h_4(k) - \frac{a_4T_s}{S}\sqrt{2gh_4(k)} + \frac{(1 - \gamma_1 - w_1(k))T_s}{3600S}q_1(k),
		\end{split}
	\end{equation}
\end{subequations}
where $T_s = 15\mathrm{s}$.
The state of the system is composed of the levels of the four tanks, which is $x \coloneqq (h_1,h_2,h_3,h_4)$, while the flows of the two pumps are the controllable inputs of the system, namely $u = (q_1,q_2)$. The parameters $\gamma_1$ and $\gamma_2$, which are characteristic parameters of the two three-way valves that act on the system, are considered to be perturbed by $w_1(k)$ and $w_2(k)$ respectively. In detail, $w(k) \coloneqq \left( w_1(k),w_2(k) \right)$ is considered to belong to a compact set, i.e.
\begin{equation*}
	w(k) \in \mathcal{W} \coloneqq \{w \in \mathbb{R}^2: \; \vert w_i \vert \leq 0.0325, \; \forall i \in \{1,2\}\}.
\end{equation*}

The parameters of the system are taken from \cite{Limon2018_NMPCT} and listed in Table \ref{tab:4tank_params} for the sake of completeness.
\begin{table}[h!]
	\centering
	\begin{tabular}{|c|c|c|}
		\hline
		\textbf{Name} & \textbf{Value} & \textbf{Description} \\ \hline
		 $a_1$ & 1.2938e-4 $m^2$ & Discharge constant of tank 1 \\ \hline
		 $a_2$ & 1.5041e-4 $m^2$ & Discharge constant of tank 2 \\ \hline
		 $a_3$ & 1.0208e-4 $m^2$ & Discharge constant of tank 3 \\ \hline
		 $a_4$ & 9.3258e-5 $m^2$ & Discharge constant of tank 4 \\ \hline
		 $S$ & 0.06 $m^2$ & Cross-section of all tanks \\ \hline
		 $\gamma_a$ & 0.3 & Parameter of the three-ways valve \\ \hline
		 $\gamma_b$ & 0.4 & Parameter of the three-ways valve \\ \hline
	\end{tabular}
	\caption{Plant parameters estimated on the real plant considered in \cite{Limon2018_NMPCT}.}
	\label{tab:4tank_params}
\end{table}
The following state and input constraints have to be verified $\forall k \geq 0$:
\begin{equation*}
	\begin{split}
		0.2\mathrm{m} \leq &h_1(k) \leq 1.36\mathrm{m},\\
		0.2\mathrm{m} \leq &h_2(k) \leq 1.36\mathrm{m},\\
		0.2\mathrm{m} \leq &h_3(k) \leq 1.30\mathrm{m},\\
		0.2\mathrm{m} \leq &h_4(k) \leq 1.30\mathrm{m},\\
		0\mathrm{m^3/h} \leq &q_1(k) \leq 3.6\mathrm{m^3/h},\\
		0\mathrm{m^3/h} \leq &q_2(k) \leq 4.0\mathrm{m^3/h}.
	\end{split}
\end{equation*}

The goal is to control the system to the equilibrium $(x_\mathrm{ref},u_\mathrm{ref})$ given by $x_\mathrm{ref} = (0.6702,0.6549,0.5435,0.5887)\mathrm{m}$ and $u_\mathrm{ref} = (1.63,2)\mathrm{m^3/h}$. The initial condition is $x(0) = (1.3533,1.1751,1.2228,0.8863)\mathrm{m}$.

Given that the aim of the current subsection is to give an insight about how to adapt standard terminal ingredients so that they can be exploited during the design of our robust contraction-based MPC, the remainder of the subsection is organized as follows: (i) we first present the sequences $\{\mathcal{F}(j)\}_{j \geq 0}$ and $\{\mathcal{R}(j)\}_{j \geq 0}$, then (ii) we introduce the workflow for computing the terminal ingredients of a standard MPC, (iii) show how such ingredients can be adapted for being used inside our robust contraction-based MPC, and (iv) list the chosen controller settings. At the end, we also show the results of our closed-loop simulations.

\subsubsection{Computation of the sequences $\{\mathcal{F}(j)\}_{j \geq 0}$ and $\{\mathcal{R}(j)\}_{j \geq 0}$}
The computation of the sequences $\{\mathcal{F}(j)\}_{j \geq 0}$ and $\{\mathcal{R}(j)\}_{j \geq 0}$ is carried out in the same way as in Section \ref{subsubsec:nhs_F_R}, as also system \eqref{eq:quadruple_tank} is component-wise Lipschitz constant. In detail, there exist non-negative real constants $L_{x,ia}$, $L_{u,ib}$ and $L_{w,ic}$, where $i,a \in \{1,2,3,4\}$, $b,c \in \{1,2\}$, such that
\begin{equation}
	\label{eq:comp_Lipschitz_4t}
	\begin{split}
		\vert f_i(x,u,w) - f_i(\check{x},\check{u},\check{w}) \vert &\leq \sum_{a = 1}^4 L_{x,ia} \vert x_a - \check{x}_a \vert + \sum_{b = 1}^2 L_{u,ib} \vert u_b - \check{u}_b \vert\\ &+ \sum_{c = 1}^2 L_{w,ic} \vert w_c - \check{w}_c \vert.
	\end{split}
\end{equation}
The values of the constants $L_{x,ia}$, $L_{u,ib}$ and $L_{w,ic}$ are respectively reported in Tables \ref{tab:cmpc_Lx_4t}, \ref{tab:cmpc_Lu_4t} and \ref{tab:cmpc_Lw_4t}.

\begin{table}[h]
	\centering
	\arrayrulecolor{black}
	\begin{tabular}{|c|c|c|c|c|c|} 
		\hline
		\rowcolor{myblue!10} \multicolumn{2}{|c|}{{\cellcolor{myblue!10}}} & \multicolumn{4}{c|}{$a$} \\ 
		\hhline{|>{\arrayrulecolor{myblue!10}}-->{\arrayrulecolor{black}}----|}
		\rowcolor{myblue!10} \multicolumn{2}{|c|}{\multirow{-2}{*}{{\cellcolor{myblue!10}}$L_{x,ia}$}} & \textit{1} & \textit{2} & \textit{3} & \textit{4} \\ 
		\hline
		{\cellcolor{myblue!10}} & {\cellcolor{myblue!10}}\textit{1} & 0.95 & 0 & 0.18 & 0 \\ 
		\hhline{|>{\arrayrulecolor{myblue!10}}->{\arrayrulecolor{black}}-----|}
		{\cellcolor{myblue!10}} & {\cellcolor{myblue!10}}\textit{2} & 0 & 0.95 & 0 & 0.15 \\ 
		\hhline{|>{\arrayrulecolor{myblue!10}}->{\arrayrulecolor{black}}-----|}
		{\cellcolor{myblue!10}} & {\cellcolor{myblue!10}}\textit{3} & 0 & 0 & 0.96 & 0 \\ 
		\hhline{|>{\arrayrulecolor{myblue!10}}->{\arrayrulecolor{black}}-----|}
		\multirow{-4}{*}{{\cellcolor{myblue!10}}$i$} & {\cellcolor{myblue!10}}\textit{4} & 0 & 0 & 0 & 0.96 \\
		\hline
	\end{tabular}
	\caption{Values of the constants $L_{x,ia}$ of the quadruple-tank system.}
	\label{tab:cmpc_Lx_4t}
\end{table}

\begin{table}[h]
	\centering
	\arrayrulecolor{black}
	\begin{tabular}{|c|c|c|c|} 
		\hline
		\rowcolor{myblue!10} \multicolumn{2}{|c|}{{\cellcolor{myblue!10}}} & \multicolumn{2}{c|}{$b$} \\ 
		\hhline{|>{\arrayrulecolor{myblue!10}}-->{\arrayrulecolor{black}}--|}
		\rowcolor{myblue!10} \multicolumn{2}{|c|}{\multirow{-2}{*}{{\cellcolor{myblue!10}}$L_{u,ib}$}} & \textit{1} & \textit{2} \\ 
		\hline
		{\cellcolor{myblue!10}} & {\cellcolor{myblue!10}}\textit{1} & 0.025 & 0 \\ 
		\hhline{|>{\arrayrulecolor{myblue!10}}->{\arrayrulecolor{black}}---|}
		{\cellcolor{myblue!10}} & {\cellcolor{myblue!10}}\textit{2} & 0 & 0.035 \\ 
		\hhline{|>{\arrayrulecolor{myblue!10}}->{\arrayrulecolor{black}}---|}
		{\cellcolor{myblue!10}} & {\cellcolor{myblue!10}}\textit{3} & 0 & 0.17 \\ 
		\hhline{|>{\arrayrulecolor{myblue!10}}->{\arrayrulecolor{black}}---|}
		\multirow{-4}{*}{{\cellcolor{myblue!10}}$i$} & {\cellcolor{myblue!10}}\textit{4} & 0.17 & 0 \\
		\hline
	\end{tabular}
	\caption{Values of the constants $L_{u,ib}$ of the quadruple-tank system.}
	\label{tab:cmpc_Lu_4t}
\end{table}

\begin{table}[h]
	\centering
	\arrayrulecolor{black}
	\begin{tabular}{|c|c|c|c|} 
		\hline
		\rowcolor{myblue!10} \multicolumn{2}{|c|}{{\cellcolor{myblue!10}}} & \multicolumn{2}{c|}{$c$} \\ 
		\hhline{|>{\arrayrulecolor{myblue!10}}-->{\arrayrulecolor{black}}--|}
		\rowcolor{myblue!10} \multicolumn{2}{|c|}{\multirow{-2}{*}{{\cellcolor{myblue!10}}$L_{w,ic}$}} & \textit{1} & \textit{2} \\ 
		\hline
		{\cellcolor{myblue!10}} & {\cellcolor{myblue!10}}\textit{1} & 0.25 & 0 \\ 
		\hhline{|>{\arrayrulecolor{myblue!10}}->{\arrayrulecolor{black}}---|}
		{\cellcolor{myblue!10}} & {\cellcolor{myblue!10}}\textit{2} & 0 & 0.275 \\ 
		\hhline{|>{\arrayrulecolor{myblue!10}}->{\arrayrulecolor{black}}---|}
		{\cellcolor{myblue!10}} & {\cellcolor{myblue!10}}\textit{3} & 0 & 0.275 \\ 
		\hhline{|>{\arrayrulecolor{myblue!10}}->{\arrayrulecolor{black}}---|}
		\multirow{-4}{*}{{\cellcolor{myblue!10}}$i$} & {\cellcolor{myblue!10}}\textit{4} & 0.25 & 0 \\
		\hline
	\end{tabular}
	\caption{Values of the constants $L_{w,ic}$ of the quadruple-tank system.}
	\label{tab:cmpc_Lw_4t}
\end{table}
The computed sequences $\{ \mathcal{F}(j) \}_{j \geq 0}$ and $\{ \mathcal{R}(j) \}_{j \geq 0}$ are such that
\begin{equation*}
	\mathcal{F}(j) = \{ x \in \mathbb{R}^4: \; \vert x_i \vert \leq \bar{x}_{i,\mathcal{F}(j)}, \forall i \in \{1,2,3,4\} \}
\end{equation*}
and
\begin{equation*}
	\mathcal{R}(j) = \{ x \in \mathbb{R}^4: \; \vert x_i \vert \leq \bar{x}_{i,\mathcal{R}(j)}, \forall i \in \{1,2,3,4\} \}.
\end{equation*}
Therefore, as for the example in Section \ref{subsec:nonholonomic_system}, we provide the computed $\bar{x}_{i,\mathcal{F}(j)}$ and $\bar{x}_{i,\mathcal{R}(j)}$ in Table \ref{tab:cmpc_F_R_values_4t}.
\begin{table}
	\centering
	\arrayrulecolor{black}
	\resizebox{\columnwidth}{!}{%
	\begin{tabular}{|c|c|c|c|c|c|c|c|c|} 
		\hline
		\rowcolor{myblue!10} {\cellcolor{myblue!10}} & \multicolumn{4}{c|}{$\mathcal{F}(j)$} & \multicolumn{4}{c|}{$\mathcal{R}(j)$} \\ 
		\hhline{|>{\arrayrulecolor{myblue!10}}->{\arrayrulecolor{black}}--------|}
		\rowcolor{myblue!10} \multirow{-2}{*}{{\cellcolor{myblue!10}}$j$} & $\bar{x}_{1,\mathcal{F}(j)}$ & $\bar{x}_{2,\mathcal{F}(j)}$ & $\bar{x}_{3,\mathcal{F}(j)}$ & $\bar{x}_{4,\mathcal{F}(j)}$ & $\bar{x}_{1,\mathcal{R}(j)}$ & $\bar{x}_{2,\mathcal{R}(j)}$ & $\bar{x}_{3,\mathcal{R}(j)}$ & $\bar{x}_{4,\mathcal{R}(j)}$ \\ 
		\hline
		{\cellcolor{myblue!10}}\textit{0} & 0.0081 & 0.0089 & 0.0089 & 0.0081 & 0 & 0 & 0 & 0 \\ 
		\hline
		{\cellcolor{myblue!10}}\textit{1} & 0.0093 & 0.0097 & 0.0086 & 0.0078 & 0.0081 & 0.0089 & 0.0089 & 0.0081 \\ 
		\hline
		{\cellcolor{myblue!10}}\textit{2} & 0.0104 & 0.0104 & 0.0082 & 0.0075 & 0.0175 & 0.0186 & 0.0175 & 0.0159 \\ 
		\hline
		{\cellcolor{myblue!10}}\textit{3} & 0.0114 & 0.0110 & 0.0079 & 0.0072 & 0.0279 & 0.0290 & 0.0258 & 0.0234 \\ 
		\hline
		{\cellcolor{myblue!10}}\textit{4} & 0.0122 & 0.0115 & 0.0076 & 0.0069 & 0.0392 & 0.0400 & 0.0337 & 0.0306 \\ 
		\hline
		{\cellcolor{myblue!10}}\textit{5} & 0.0130 & 0.0120 & 0.0073 & 0.0066 & 0.0514 & 0.0516 & 0.0413 & 0.0375 \\ 
		\hline
		{\cellcolor{myblue!10}}\textit{6} & 0.0136 & 0.0124 & 0.0070 & 0.0064 & 0.0644 & 0.0635 & 0.0485 & 0.0441 \\ 
		\hline
		{\cellcolor{myblue!10}}\textit{7} & 0.0142 & 0.0127 & 0.0067 & 0.0061 & 0.0781 & 0.0759 & 0.0555 & 0.0505 \\ 
		\hline
		{\cellcolor{myblue!10}}\textit{8} & 0.0147 & 0.0130 & 0.0064 & 0.0059 & 0.0923 & 0.0886 & 0.0623 & 0.0566 \\ 
		\hline
		{\cellcolor{myblue!10}}\textit{9} & 0.0151 & 0.0132 & 0.0062 & 0.0056 & 0.1070 & 0.1016 & 0.0687 & 0.0625 \\ 
		\hline
		{\cellcolor{myblue!10}}\textit{10} & 0.0155 & 0.0134 & 0.0059 & 0.0054 & 0.1221 & 0.1149 & 0.0749 & 0.0681 \\ 
		\hline
		{\cellcolor{myblue!10}}\textit{11} & 0.0158 & 0.0135 & 0.0057 & 0.0052 & 0.1376 & 0.1283 & 0.0808 & 0.0735 \\ 
		\hline
		{\cellcolor{myblue!10}}\textit{12} & 0.0160 & 0.0136 & 0.0055 & 0.0050 & 0.1534 & 0.1418 & 0.0865 & 0.0787 \\ 
		\hline
		{\cellcolor{myblue!10}}\textit{13} & 0.0162 & 0.0137 & 0.0053 & 0.0048 & 0.1695 & 0.1555 & 0.0920 & 0.0836 \\ 
		\hline
		{\cellcolor{myblue!10}}\textit{14} & 0.0163 & 0.0137 & 0.0050 & 0.0046 & 0.1857 & 0.1692 & 0.0973 & 0.0884 \\ 
		\hline
		{\cellcolor{myblue!10}}\textit{15} & 0.0164 & 0.0137 & 0.0048 & 0.0044 & 0.2020 & 0.1829 & 0.1023 & 0.0930 \\ 
		\hline
		{\cellcolor{myblue!10}}\textit{16} & 0.0165 & 0.0137 & 0.0047 & 0.0042 & 0.2185 & 0.1967 & 0.1072 & 0.0974 \\ 
		\hline
		{\cellcolor{myblue!10}}\textit{17} & 0.0165 & 0.0137 & 0.0045 & 0.0041 & 0.2350 & 0.2104 & 0.1118 & 0.1016 \\
		\hline
	\end{tabular}%
	}
	\caption{Description of the sequences $\{ \mathcal{F}(j) \}_{j \geq 0}$ and $\{ \mathcal{R}(j) \}_{j \geq 0}$ related to the quadruple-tank system that were obtained by means of the algorithm shown in Section \ref{subsec:F_R_Lipschitz}. The values for $j > 17$ are not reported because, as explained later in the text, the maximum prediction horizon of our MPC is set to $N_p = 17$.}
	\label{tab:cmpc_F_R_values_4t}
\end{table}

\subsubsection{Computation of the terminal ingredients}
\label{subsubsec:terminal_ingredients}
In order to introduce the design of the terminal ingredients of a standard MPC with terminal cost and terminal set, it is necessary to anticipate some design choices that will also be explained in Section \ref{subsubsec:controller_design_4t}. Indeed, in order to design such ingredients, the knowledge of the stage cost function $\ell(x,u)$ is needed. For this reason, we choose to employ a quadratic stage cost function
\begin{equation*}
	\ell(x,u) = (x-x_\mathrm{ref})^\top Q (x-x_\mathrm{ref}) + (u-u_\mathrm{ref})^\top R (u-u_\mathrm{ref}),
\end{equation*}
where $Q = I_4$ and $R = 0.01I_2$.

Thereafter, we need to design a terminal cost function $V_f(x)$ that behaves as a CLF inside an invariant set $\mathcal{X}_f$ thanks to a locally stabilizing control law $u = \kappa_f(x)$. In order to simplify the design process, we decide to employ a quadratic terminal cost function $V_f(x) = (x - x_\mathrm{ref})^\top P (x - x_\mathrm{ref})$, where $P$ is simmetric and positive definite, to use as terminal set a sublevel set of $V_f(x)$, namely $\mathcal{X}_f \coloneqq \{ x \in \mathbb{R}^4: \; V_f(x) \leq \beta \}$ for some $\beta > 0$, and to employ a locally stabilizing linear controller, i.e. $u = u_\mathrm{ref} + K(x - x_\mathrm{ref})$, for some $K$. In particular, as often done in the literature (e.g. \cite{rawlings2017model}, \cite{allgower1998qihmpc}, \cite{lazar2018terminal}), we assume to design these ingredients starting from the linearization of system \eqref{eq:quadruple_tank} at $x_\mathrm{ref}$. 

The linearized system is described by the matrices
\begin{equation*}
	A = \begin{bmatrix}
		0.9125 & 0 & 0.0767 & 0 \\
		0 & 0.8971 & 0 & 0.0673 \\
		0 & 0 & 0.9233 & 0 \\
		0 & 0 & 0 & 0.9327
	\end{bmatrix}
\end{equation*}
and 
\begin{equation*}
	B = \begin{bmatrix}
		0.0208 & 0 \\ 0 & 0.0278 \\ 0 & 0.0417 \\ 0.0486 & 0
	\end{bmatrix}.
\end{equation*}
Provided that the couple $(A,B)$ is fully controllable, it is possible to compute $P$, $K$, and $\beta$ by following standard methods based on $A$ and $B$. 

The theoretical details behind the computation of such ingredients are reported in the Appendix \ref{subsec:algorithm_terminal_ingredients}. For our current goal, it is sufficient to report the LMI that can be solved to find their values. In detail, given $\kappa \in (0,1)$ and setting $S = P^{-1}$ and $O = KP^{-1}$, the following LMI has to be solved:
\begin{equation*}
	\begin{bmatrix}
		(1 - \kappa)S & (AS+BO)^\top & S & O^\top \\
		AS+BO & S & 0 & 0 \\
		S & 0 & Q^{-1} & 0 \\
		O & 0 & 0 & R^{-1}
	\end{bmatrix} \geq 0.
\end{equation*}
Then, given $P = S^{-1}$ and $K = OP$, the following conditions have to be verified:
\begin{subequations}
	\begin{equation*}
		\begin{split}
			e(x)^\top P e(x) &+ 2e(x)^\top P(A+BK)(x - x_\mathrm{ref})\\ &\leq \kappa (x - x_\mathrm{ref})^\top P (x - x_\mathrm{ref}), \; \forall x \; \mathrm{s.t.} \; V_f(x) \leq \beta,
		\end{split}
	\end{equation*}
	\begin{equation*}
		u_\mathrm{ref} + K(x - x_\mathrm{ref}) \in \mathcal{U}, \; \forall x \; \mathrm{s.t.} \; V_f(x) \leq \beta,
	\end{equation*}
\end{subequations}
where $e(x)$ is defined as $e(x) = f(x,u_\mathrm{ref}+K(x-x_\mathrm{ref}),0) - \left(x_\mathrm{ref} + (A+BK)(x - x_\mathrm{ref}) \right)$. 

Setting $\kappa = 0.025$, by following this method we get:
\begin{equation*}
	P = \begin{bmatrix}
		6.0794 & -0.9107 & 1.5580 & -1.9296 \\
		-0.9107 & 4.9770 & -2.1981 & 1.0145 \\
		1.5580 & -2.1981 & 4.1999 & -1.0133 \\
		-1.9296 & 1.0145 & -1.0133 & 3.3115
	\end{bmatrix},
\end{equation*}
\begin{equation*}
	K = \begin{bmatrix}
		-1.7914 & -1.6219 & 0.8432 & -6.9335 \\
		-2.2376 & -2.5948 & -6.7541 & 0.6823
	\end{bmatrix},
\end{equation*}
\begin{equation*}
	\beta = 0.124.
\end{equation*}

\subsubsection{Adaptation of the terminal ingredients for usage in our robust contraction-based MPC}
In order to recycle the terminal ingredients introduced earlier, the first thing to do is check whether the invariant set $\mathcal{X}_f$ is also robustly invariant. Indeed, the set $\mathcal{X}_f$ being invaraint for the nominal system does not ensure that the invariance property is kept also for all the possible perturbation realizations $w \in \mathcal{W}$. In our case, $\mathcal{X}_f$ is also robustly invariant, therefore we can set $\Omega = \mathcal{X}_f$. 

Moreover, note that $V_f(x)$ is positive definite by design and that the interior of the set $\Omega \ominus \mathcal{R}(1)$ is non-empty. For this reason, we can amploy $V_f(x)$ as contractive function, i.e. $\Gamma(x) = V_f(x)$, and we can be sure that there exists a $\omega > 0$ such that $\{x: \; \Gamma(x) \leq \omega\} \subseteq \Omega \ominus \mathcal{R}(1)$. In our case, $\omega = 0.074$.

We now proceed by computing the upper bound on the contractivity factor $\gamma$, which is computed as $\omega/\Gamma_\mathrm{max}$. Given that $\Gamma_\mathrm{max} = 7.7408$, we get that the contraction factor must be such that $\gamma \leq 0.0096$.
\begin{remark}
	The found upper bound on $\gamma$ clearly shows how recycling the terminal ingredients used in a standard MPC is possible but inefficient. Indeed, imposing such a low contractivity factor is very similar to imposing a terminal equality constraint, consequently leading to large prediction horizons.
\end{remark} 
The same algorithm shown in Section \ref{subsubsec:contractive_function} is used to compute the lowest maximum prediction horizon $N_p$ that ensures the needed contractivity and the corresponding $\gamma$ value. The obtained results are $N_p = 17$ and $\gamma = 0.0079$, which verify all the conditions and ensure that the the sets $\mathcal{X} \ominus \mathcal{R}(j)$, $\forall j \in [1,N_p]$, are non-empty.

\subsubsection{Controller settings}
\label{subsubsec:controller_design_4t}
The employed stage cost function has already been anticipated in Section \ref{subsubsec:terminal_ingredients}. In order for the paramater $\xi$ to verify all the conditions of Theorem \ref{thm:cmpc_2op}, its value is set to $\xi = 73.0013$, as the maximum value of the stage cost function over the set $\mathcal{X} \times \mathcal{U}$ is $\bar{\ell} = 2.13$.

Regarding the dynamics of the internal controller state $\theta$, its update parameters $\nu$ and $\epsilon$ are set to $\nu = 0.99$ and $\epsilon = 10^{-8}$, while its initial condition is set to $\theta(0) = \nu \Gamma \left( x(0) \right) = 4.7323$.

\subsubsection{Simulation results}
As also done for the example in Section \ref{subsec:nonholonomic_system}, we run 100 simulations of the closed-loop system, setting the length of each simulation to 50 steps, i.e. 750s. All the simulations start from the given initial condition and use a perturbation $w$ that is randomly sampled from $\mathcal{W}$ at each instant $k$.

\begin{figure}[h]
	\includegraphics[trim=3cm 7cm 3cm 7cm, clip,width=\columnwidth]{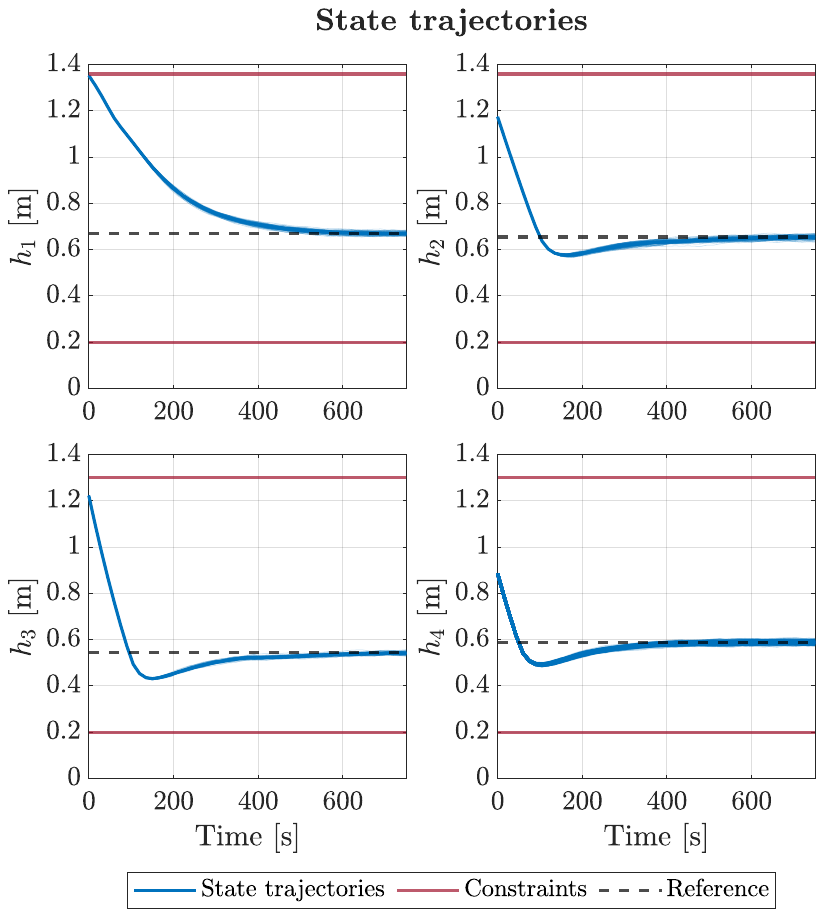}
	\centering
	\caption{Closed-loop state trajectories under our robust contraction-based MPC.} 
	\label{fig:states_4t}
\end{figure}

\begin{figure}[h]
	\includegraphics[trim=3cm 9.75cm 3cm 9cm, clip,width=\columnwidth]{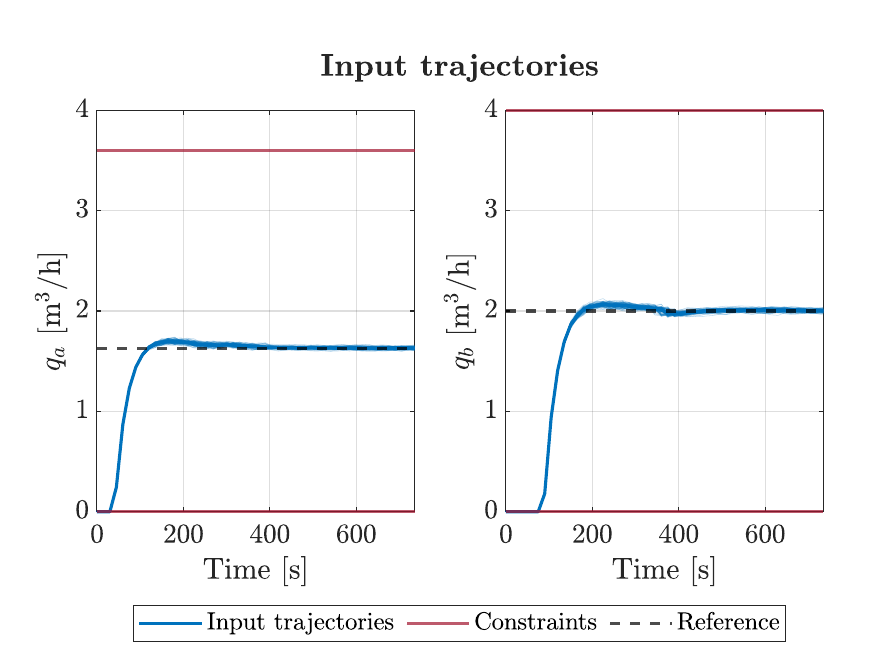}
	\centering
	\caption{Control actions provided by our robust contraction-based MPC.} 
	\label{fig:inputs_4t}
\end{figure}

\begin{figure}[h]
	\includegraphics[trim=3cm 10cm 3cm 9cm, clip,width=\columnwidth]{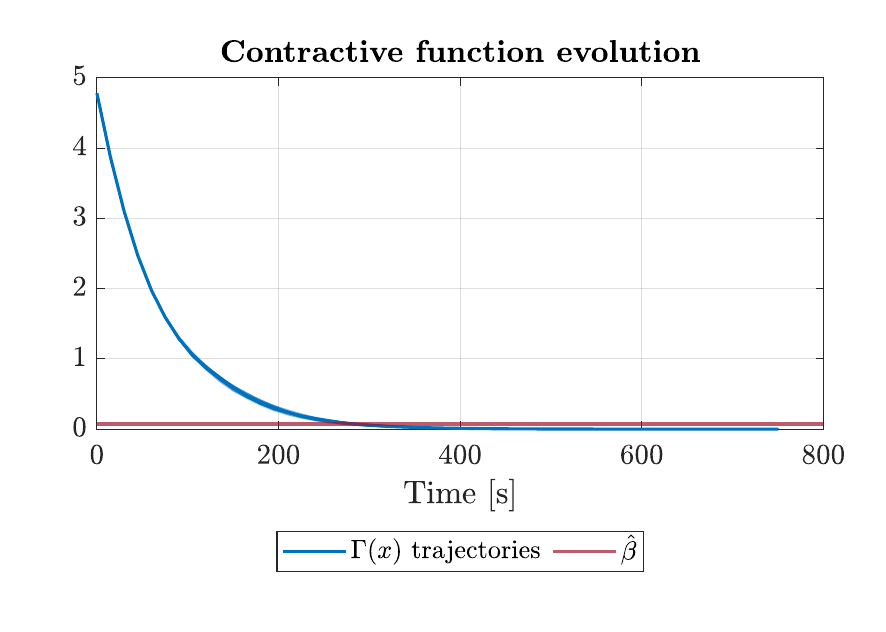}
	\centering
	\caption{Evolution of the contractive function $\Gamma(x)$ during the simulations.} 
	\label{fig:Gamma_4t}
\end{figure}

The trajectories of the system states and the control actions that are computed by our MPC at each sampling instant are respectively shown in Figure \ref{fig:states_4t} and Figure \ref{fig:inputs_4t}.
The evolution of the contractive function $\Gamma(x)$, which is shown in Figure \ref{fig:Gamma_4t}, is totally different from the one obtained on the nonholonomic system in Section \ref{subsec:nonholonomic_system}. In particular, while also in this case $\Gamma(x)$ dicreases over time and approaches zero, the value $\omega$ is reached after a relatively large number of sampling instants. This is consistent with the design process that has been followed. Indeed, for the nonholonomic system in Section \ref{subsec:nonholonomic_system}, the state constraint set $\mathcal{X}$ coincides with the RCIS $\Omega$ and the contractive function $\Gamma(x)$ is designed with the aim of guaranteeing contractivity within shorten horizons. In the case of the currently analyzed four-tank system, the contractivity ingredients are not designed with the same goal, therefore guaranteeing all the stability and robustness properties esured by our robust contraction-based MPC in a less efficient way.

\section{Conclusion}
We have introduced a new robust contraction-based MPC that is able to robustly control and stabilize perturbed nonlinear systems that meet standard conditions. The main advantages of our formulation are that no terminal constraints have to be imposed and that the terminal cost function to be designed does not need to locally behave as a CLF. Moreover, the employed prediction horizon, which is also optimized online, is potentially shorter than the one needed to guarantee the feasibility of a terminal equality constraint MPC. This can prove particularly useful when the robustness of the controller is guaranteed by imposing tightened state constraints, as longer prediction horizons could lead to empty feasible sets.

The theoretical concepts introduced in \cite{alamir2017contraction} and \cite{Limon2009iss} have been exploited to prove the stability and recursive feasibility properties of our MPC. Moreover, the examples presented in Section \ref{sec:case_studies} show the effectiveness of our MPC in controlling systems for which the design of standard terminal ingredients would be challenging, and also how terminal ingredients designed following standard methods can be recycled in our formulation.

\appendix
\section{Proofs}
\label{subsec:proofs}
\subsection{Proof of Lemma \ref{lem:cmpc_V_{N_p}_star_limited}}
Given that $P_{N_p} \left( x(k),\theta(k) \right)$ is feasible, the stage cost is surely lower than $N_p\bar{\ell}$ thanks to Assumption \ref{ass:cmpc_stage_cost_L}, while $\underline{\Gamma} \big( x(k),\mathbf{u}(k),N_p \big) \leq \gamma \Gamma \big( x(k) \big)$ as a result of Assumption \ref{ass:cmpc_contractivity}. This proves \eqref{eq:cmpc_V_{N_p}_star_limit}.

Concerning \eqref{eq:cmpc_i_star_equals_q_star}, it can be proved by contradiction. Suppose $j^* \big( x(k),\theta(k) \big) < q^* \big( x(k),\theta(k) \big)$; then, the candidate solution $\big( \mathbf{u}^c(k),q^c(k) \big) \coloneqq \big( \big\{ u^*(0 \vert k), \hdots, u^*(j^*-1 \vert k) \big\}, j^* \big( x(k), \allowbreak \theta(k) \big) \big)$, together with the related state trajectory $\hat{\mathbf{x}}^c(k) \coloneqq \{ \hat{x}^c(0 \vert k), \hdots, \allowbreak \hat{x}^c(q^c \vert k) \}$, where $\hat{x}^c(j \vert k) \coloneqq \phi(j;x(k),\mathbf{u}^c(k),\mathbf{0})$, would correspond to a cost function value satisfying
\begin{equation*}
	\begin{split}
		V_{N_p} \left( x(k),\theta(k), \mathbf{u}^c(k),q^c(k) \right) &= \theta(k) \cdot \sum_{j=0}^{q^c(k)-1} \ell \left( \hat{x}^c(j \vert k), u^c(j \vert k) \right)\\ &+ \xi \min_{j = 1}^{q^c(k)} \Gamma \left( \hat{x}^c(j \vert k) \right).
	\end{split}
\end{equation*}
Considering that
\begin{equation*}
	\begin{split}
		V_{N_p}^* \left( x(k),\theta(k) \right) &= \theta(k)\cdot \sum_{j=0}^{q^*(k)-1} \ell \left( \hat{x}^*(j \vert k), u^*(j \vert k) \right)\\ &+ \xi \min_{j = 1}^{q^*(k)} \Gamma \left( \hat{x}^*(j \vert k) \right),
	\end{split}
\end{equation*}
by optimality we get
\begin{equation}
	\label{eq:cmpc_horizon_optimality}
	\begin{split}
		0 &\leq V_{N_p} \left( x(k),\theta(k), \mathbf{u}^c(k),q^c(k) \right) - V_{N_p}^* \left( x(k),\theta(k) \right) \\
		&= \theta(k) \cdot \sum_{j=0}^{q^c(k)-1} \ell \left( \hat{x}^c(j \vert k), u^c(j \vert k) \right) + \xi \min_{j = 1}^{q^c(k)} \Gamma \left( \hat{x}^c(j \vert k) \right) \\
		&- \theta(k) \cdot \sum_{j=0}^{q^*(k)-1} \ell \left( \hat{x}^*(j \vert k), u^*(j \vert k) \right) - \xi \min_{j = 1}^{q^*(k)} \Gamma \left( \hat{x}^*(j \vert k) \right).
	\end{split}
\end{equation}
Provided that the instant of maximum contraction is $j^* \equiv q^c(k)$, we get that $$\min_{j = 1}^{q^c(k)} \Gamma \left( \hat{x}^c(j \vert k) \right) = \min_{j = 1}^{q^*(k)} \Gamma \left( \hat{x}^*(j \vert k) \right).$$ Therefore, \eqref{eq:cmpc_horizon_optimality} reduces to
\begin{equation*}
	\begin{split}
		0 &\leq V_{N_p} \left( x(k),\theta(k), \mathbf{u}^c(k),q^c(k) \right) - V_{N_p}^* \left( x(k),\theta(k) \right)\\ &= -\theta(k) \cdot \sum_{j = q^c(k)}^{q^*(k) - 1} \ell \left( \hat{x}^*(j \vert k), u^*(j \vert k) \right) \leq 0,
	\end{split}
\end{equation*}
which implies
\begin{equation*}
	\sum_{j = q^c(k)}^{q^*(k) - 1} \ell \left( \hat{x}^*(j \vert k), u^*(j \vert k) \right) = 0,
\end{equation*}
meaning that the desired equilibrium is reached after $q^c(k)$ prediction steps. This contradicts the optimality of $q^*(k)$.

\subsection{Proof of Lemma \ref{lem:cmpc_V_{N_p}_star_ineq}}
The conditions of Lemma \ref{lem:cmpc_V_{N_p}_star_limited} are satisfied. Therefore,
\begin{equation*}
	V_{N_p}^* \left( x(k),\theta(k) \right) \leq \theta(k)N_p\bar{\ell} + \xi \gamma \Gamma \left( x(k) \right).
\end{equation*}
To continue the proof, given that the update law $f_\theta( x(k),\theta(k-1) )$ ensures that $\Gamma\left( x(k) \right) > \theta(k)$ for all $x \in \mathcal{X}$ such that $\Gamma(x) > \epsilon$, we need to discern two cases:

\textbf{Case $\Gamma \left( x(k) \right) > \theta(k)$:}
Given that also $\Gamma \left( x(k) \right) > \theta(k)$ is satisfied, we get that
\begin{equation*}
	V_{N_p}^* \left( x(k),\theta(k) \right) \leq (N_p\bar{\ell} + \xi \gamma) \Gamma \left( x(k) \right).
\end{equation*}
Using the assumption $\xi \geq 2N_p\bar{\ell} / (1-\gamma)$, we obtain 
\begin{equation}
	\label{eq:cmpc_V_{N_p}_star_upper_bound_case1}
	V_{N_p}^* \left( x(k),\theta(k) \right) \leq \left[ \frac{\gamma + 1}{2} \right] \xi \Gamma \left( x(k) \right).
\end{equation}

\textbf{Case $\Gamma \left( x(k) \right) \leq \theta(k) = \epsilon$:}
Given that $\theta(k) = \epsilon$, \eqref{eq:cmpc_V_{N_p}_star_limit} rewrites as
\begin{equation}
	\label{eq:cmpc_V_{N_p}_star_upper_bound_case2}
	V_{N_p}^* \left( x(k),\theta(k) \right) \leq \epsilon N_p \bar{\ell} + \gamma \xi \Gamma \left( x(k) \right).
\end{equation}
Unifying \eqref{eq:cmpc_V_{N_p}_star_upper_bound_case1} and \eqref{eq:cmpc_V_{N_p}_star_upper_bound_case2}, we obtain
\begin{equation*}
	\begin{split}
		V_{N_p}^* \left( x(k),\theta(k) \right) &\leq \epsilon N_p \bar{\ell} + \xi \max \left\{ \gamma, \frac{\gamma + 1}{2} \right\} \Gamma \left( x(k) \right)\\
		&\epsilon N_p \bar{\ell} + \xi \left[ \frac{\gamma + 1}{2} \right] \Gamma \left( x(k) \right),
	\end{split} 
\end{equation*}
which ends the proof.

\subsection{Proof of Lemma \ref{lem:cmpc_q_greater_1}}
\textbf{Recursive feasibility:} Given that $q^*(k) > 1$, $\vert x_i(k+1) - \hat{x}^*_i(1 \vert k)\vert \leq \bar{w}_i$, $\forall i \in [1,n]$, that $u^+(j - 1 \vert k+1) = u^*(j \vert k)$, and that Assumptions \ref{ass:cmpc_F_sequence} and \ref{ass:cmpc_R_sequence} hold, we get that, $\forall j \in [1,q^+(k+1)]$,
\begin{equation*}
	\begin{split}
		\phi \left( j;x(k+1),\mathbf{u}^+(k+1),\mathbf{0} \right) &\in \phi \left( j;\hat{x}^*(1 \vert k),\mathbf{u}^+(k+1),\mathbf{0} \right) \oplus \mathcal{F}(j) \\
		&\subseteq \phi \left( j;\hat{x}^*(1 \vert k),\mathbf{u}^+(k+1),\mathbf{0} \right)\\ &\oplus \mathcal{R}(j+1) \ominus \mathcal{R}(j).
	\end{split}
\end{equation*}
As $\phi \left( j;\hat{x}^*(1 \vert k),\mathbf{u}^+(k+1),\mathbf{0} \right) \in \mathcal{X} \ominus \mathcal{R}(j+1)$, we eventually obtain that
\begin{equation*}
	\begin{split}
		\phi \left( j;x(k+1),\mathbf{u}^+(k+1),\mathbf{0} \right) &\in \phi \left( j;\hat{x}^*(1 \vert k),\mathbf{u}^+(k+1),\mathbf{0} \right)\\ &\oplus \mathcal{R}(j+1) \ominus \mathcal{R}(j) \\
		&\subseteq \left( \mathcal{X} \ominus \mathcal{R}(j+1) \right) \oplus \mathcal{R}(j+1) \ominus \mathcal{R}(j) \\
		&\subseteq \mathcal{X} \ominus \mathcal{R}(j),
	\end{split}
\end{equation*}
which proves that the couple $(\mathbf{u}^+(k+1),q^+(k+1))$ is feasible.

\textbf{Cost descent:} The cost related to the admissible pair $(\mathbf{u}^+(k+1),q^+(k+1))$ is given by
\begin{equation*}
	\begin{split}
		&V_{N_p} \left( x(k+1),\theta(k+1),\mathbf{u}^+(k+1),q^+(k+1) \right)\\ &= \theta(k+1) \cdot \sum_{j = 0}^{q^+(k+1)-1} \ell \left( \hat{x}^+(j \vert k+1), u^+(j \vert k+1) \right)\\ &+ \xi \underline{\Gamma} \left( x(k+1),\mathbf{u}^+(k+1),q^+(k+1) \right).
	\end{split}
\end{equation*}
Since $\underline{\Gamma} \left( x(k+1),\mathbf{u}^+(k+1),q^+(k+1) \right) \leq \Gamma \left( \hat{x}^+(j \vert k+1) \right)$ for all $j \in [1,q^+(k+1)]$, we get:
\begin{equation*}
	\begin{split}
		&V_{N_p} \left( x(k+1),\theta(k+1),\mathbf{u}^+(k+1),q^+(k+1) \right)\\ &\leq \theta(k+1) \cdot \sum_{j = 0}^{q^+(k+1)-1} \ell \left( \hat{x}^+(j \vert k+1), u^+(j \vert k+1) \right)\\ &+ \xi \Gamma \left( \hat{x}^+(q^+ \vert k+1) \right).
	\end{split}
\end{equation*}
Let us now evaluate $V_{N_p} \left( x(k+1),\theta(k+1),\mathbf{u}^+(k+1),q^+(k+1) \right) - V_{N_p}^* \left( x(k),\theta(k) \right)$. Taking into account that, in virtue of Lemma \ref{lem:cmpc_V_{N_p}_star_ineq}, $q^*(x(k),\theta(k)) = j^*(x(k),\theta(k))$, we also have that $\underline{\Gamma} \left( x(k),\mathbf{u}^*(k),q^*(k) \right) = \Gamma \left( \hat{x}^*(q^*(k) \vert k) \right)$, therefore:
\begin{equation*}
	\begin{split}
		&V_{N_p} \left( x(k+1),\theta(k+1),\mathbf{u}^+(k+1),q^+(k+1) \right) - V_{N_p}^* \left( x(k),\theta(k) \right)\\ &\leq \theta(k+1) \cdot \sum_{j = 0}^{q^+(k+1)-1} \ell \left( \hat{x}^+(j \vert k+1), u^+(j \vert k+1) \right)\\ &+ \xi \Gamma \left( \hat{x}^+(q^+(k+1) \vert k+1) \right) \\
		&- \theta(k) \cdot \sum_{j = 0}^{q^*(k)-1} \ell \left( \hat{x}^*(j \vert k), u^*(j \vert k) \right) - \xi \Gamma \left( \hat{x}^*(q^*(k) \vert k) \right).
	\end{split}		 
\end{equation*}

Since $\theta(k+1) = f_\theta \left( x(k+1),\theta(k) \right)$, we have that $\theta(k+1) \leq \theta(k)$; for this reason, we obtain
\begin{equation*}
	\begin{split}
		&V_{N_p} \big( x(k+1),\theta(k+1),\mathbf{u}^+(k+1),q^+(k+1) \big) - V_{N_p}^* \big( x(k),\theta(k) \big)\\ &\leq \theta(k) \cdot \sum_{j = 0}^{q^+(k+1)-1} \ell \big( \hat{x}^+(j \vert k+1), u^+(j \vert k+1) \big)\\ &+ \xi \Gamma \big( \hat{x}^+(q^+(k+1) \vert k+1) \big) \\
		&- \theta(k) \cdot \sum_{j = 0}^{q^*(k)-1} \ell \left( \hat{x}^*(j \vert k), u^*(j \vert k) \right) - \xi \Gamma \left( \hat{x}^*(q^*(k) \vert k) \right) \\
		&= -\theta(k) \ell \left( x(k),u^*(k) \right)   +   \xi \Big[ \Gamma \left( \hat{x}^+(q^+(k+1) \vert k   +   1) \right)\\
		&-   \Gamma \left( \hat{x}^*(q^*(k) \vert k) \right) \Big]\\ &+ \theta(k) \cdot \sum_{j = 1}^{q^*(k)-1} \big[ \ell \left( \hat{x}^+(j-1 \vert k   +   1), u^+(j-1 \vert k+1) \right)\\ &- \ell \big( \hat{x}^*(j \vert k), u^*(j \vert k) \big) \big].
	\end{split}		 
\end{equation*}
Because of Assumptions \ref{ass:cmpc_contractivity} and \ref{ass:cmpc_stage_cost_L}, and due to the fact that $u^+(j-1 \vert k+1) \equiv u^*(j \vert k)$, the last inequality becomes
\begin{equation*}
	\begin{split}
		&V_{N_p} \left( x(k+1),\theta(k+1),\mathbf{u}^+,q^+ \right) - V_{N_p}^* \left( x(k),\theta(k) \right)\\ &\leq -\theta(k) \alpha_\ell \left( \Vert x(k) \Vert \right) + \xi \sigma_\Gamma \left( \left\Vert \hat{x}^+(q^*-1 \vert k+1) - \hat{x}^*(q^* \vert k) \right\Vert \right) \\
		&+ \theta(k) \cdot \sum_{j = 1}^{q^*-1} \sigma_{\ell,x} \left( \left\Vert \hat{x}^+(j-1 \vert k+1) - \hat{x}^*(j \vert k) \right\Vert \right).
	\end{split}
\end{equation*}

Let us analyze the term $\left\Vert \hat{x}^+(j-1 \vert k+1) - \hat{x}^*(j \vert k) \right\Vert$. In particular, note that
\begin{equation*}
	\Vert \hat{x}^+(j  - 1 \vert k  + 1)  -  \hat{x}^*(j \vert k) \Vert = \begin{Vmatrix}
		\hat{x}^+_1(j  - 1 \vert k  + 1)  -  \hat{x}^*_1(j \vert k) \\
		\vdots \\
		\hat{x}^+_n(j  - 1 \vert k  + 1)  -  \hat{x}^*_n (j \vert k)
	\end{Vmatrix}.
\end{equation*}
From the component-wise uniform continuity of the model, we have that, for all $i \in [1,n]$,
\begin{equation*} 
	\vert \hat{x}_i^+(0 \vert k+1) - \hat{x}_i^*(1 \vert k) \vert \leq \sum_{a = 1}^r \sigma_{w,ia} (w_a(k)) \eqqcolon c_{i,0}(w(k)).
\end{equation*}
In the same way,
\begin{equation*} 
	\begin{split}
		\vert \hat{x}_i^+(1 \vert k+1) - \hat{x}_i^*(2 \vert k) \vert &\leq \sum_{b = 1}^n \sigma_{x,ib} \left( \sum_{a = 1}^r \sigma_{w,ba} (w_a(k)) \right)\\ &= \sum_{b = 1}^n \sigma_{x,ib} (c_{b,0}(w(k))) \eqqcolon c_{i,1}(w(k)).
	\end{split}
\end{equation*}
Generalizing for all $j > 0$,
\begin{equation*} 
	\begin{split}
		\vert \hat{x}_i^+(j \vert k+1) - \hat{x}_i^*(j+1 \vert k) \vert &\leq \sum_{b = 1}^n \sigma_{x,ib} \left( c_{b,j-1}(w(k)) \right) \eqqcolon c_{i,j}(w(k)).
	\end{split}
\end{equation*}
Therefore,
\begin{equation}
	\label{eq:cmpc_r_j}
	\begin{split}
		\Vert \hat{x}^+(j \vert k + 1) - \hat{x}^*(j+1 \vert k) \Vert &\leq \begin{Vmatrix}
			c_{1,j}(w(k)) \\
			\vdots \\
			c_{n,j}(w(k))
		\end{Vmatrix} \eqqcolon r_j(w(k)).
	\end{split}
\end{equation}
Note that the functions $r_j(w(k))$ are positive definite,  strictly increasing with respect to $\Vert w(k) \Vert$, and they tend to infinity for $\Vert w(k) \Vert \to \infty$; for this reason, we can conclude that they are $\mathcal{K}_\infty$ functions.

Consequently, we obtain
\begin{equation*}
	\begin{split}
		&V_{N_p} \left( x(k+1),\theta(k+1),\mathbf{u}^+(k+1),q^+(k+1) \right) - V_{N_p}^* \left( x(k),\theta(k) \right)\\ &\leq -\theta(k) \alpha_\ell \left( \Vert x(k) \Vert \right) + \xi \sigma_\Gamma \left( r_{q^*(k)-1} \left( w(k) \right) \right)\\
		&+ \theta(k) \cdot \sum_{j = 1}^{q^*(k)-1} \sigma_{\ell,x} \left( r_{j-1} \left( w(k) \right) \right).
	\end{split}
\end{equation*}
Provided that there exists a $\mathcal{K}$ function $\lambda_1(\cdot)$ such that the following conditions hold simultaneously:
\begin{enumerate}
	\item $\lambda_1(\Vert w \Vert) \geq \sum_{j = 1}^{q^*(k)-1} \sigma_{\ell,x} \left( r_{j-1} \left( w \right) \right)$,
	\item $\lambda_1(\Vert w \Vert) \geq \sigma_\Gamma \left( r_{q^*(k)-1} \left( w \right) \right)$,
\end{enumerate}
the last inequality becomes
\begin{equation*}
	\begin{split}
		&V_{N_p} \left( x(k+1),\theta(k+1),\mathbf{u}^+(k+1),q^+(k+1) \right) - V_{N_p}^* \left( x(k),\theta(k) \right)\\ &\leq -\theta(k) \alpha_\ell \left( \Vert x(k) \Vert \right) + \left( \theta(k)+\xi \right) \lambda_1 \left( \Vert w(k) \Vert \right).
	\end{split}
\end{equation*}

\subsection{Proof of Lemma \ref{lem:cmpc_q_equal_1}}
\textbf{Recursive feasibility:} Given that $q^* = 1$, we have that $$\Gamma(\hat{x}^*(1 \vert k)) \leq \gamma \Gamma(x(k)) \leq \omega \frac{\Gamma(x(k))}{\Gamma_\mathrm{max}} \leq \omega.$$ Therefore, $\hat{x}^*(1 \vert k) \in \Omega \ominus \mathcal{R}(1)$, which implies $x(k+1) \in \hat{x}^*(1 \vert k) \oplus \mathcal{R}(1) \subset \Omega$.

Since, thanks to Assumption \ref{ass:cmpc_rcis}, $\Omega \subseteq \mathcal{X} \cap \mathcal{Q} \left( \mathcal{X} \ominus \mathcal{R}(1) \right)$ for the nominal system, there exists $u^+(0 \vert k+1) \in \mathcal{U}$ that is able to drive the nominal system state to $\mathcal{X} \ominus \mathcal{R}(1)$. Therefore, a feasible solution to $P_{N_p}(x(k+1),\theta(k+1))$ is $(q^+(k+1),\mathbf{u}^+(k+1)) = (1,\{ u^+(0 \vert k+1) \})$.

\textbf{Cost descent:} Since $q^*(k) = 1$ and by virtue of the receding horizon implementation, we have
\begin{equation*}
	\begin{split}
		V_{N_p}^* \left( x(k),\theta(k) \right) &= \theta(k) \ell \left( x(k),u^*(k) \right) + \xi \Gamma \left( \hat{x}^*(1 \vert k) \right)\\ &\geq \theta(k) \alpha_\ell \left( \Vert x(k) \Vert \right) + \xi \Gamma \left( \hat{x}^*(1 \vert k) \right).
	\end{split}
\end{equation*}
Since the conditions of Lemma \ref{lem:cmpc_V_{N_p}_star_ineq} are satisfied at time instant $k+1$, we have 
\begin{equation*}
	\begin{split}
		V_{N_p}^* \left( x(k+1),\theta(k+1) \right) &\leq \left[ \frac{1+\gamma}{2} \right] \xi \Gamma \left( x(k+1) \right) + \epsilon N_p \bar{\ell}\\ &< \xi \Gamma \left( x(k+1) \right) + \epsilon N_p \bar{\ell}.
	\end{split}
\end{equation*}
Putting together the last two inequalities, we get
\begin{equation*}
	\begin{split}
		&V_{N_p}^* \left( x(k+1),\theta(k+1) \right) - V_{N_p}^* \left( x(k),\theta(k) \right)\\ &\leq \xi \Gamma \left( x(k+1) \right) - \theta(k) \alpha_\ell \left( \Vert x(k) \Vert \right) - \xi \Gamma \left( \hat{x}^*(1 \vert k) \right) + \epsilon N_p \bar{\ell}.
	\end{split}
\end{equation*}
Due to the uniform continuity of $\Gamma(\cdot)$ over $\mathcal{X}$, this becomes:
\begin{equation*}
	\begin{split}
		&V_{N_p}^* \left( x(k+1),\theta(k+1) \right) - V_{N_p}^* \left( x(k),\theta(k) \right)\\ &\leq - \theta(k) \alpha_\ell \left( \Vert x(k) \Vert \right) + \xi \sigma_\Gamma \left( \left\Vert x(k+1) - \hat{x}^*(1 \vert k) \right\Vert \right) + \epsilon N_p \bar{\ell}.
	\end{split}
\end{equation*}
Given that, as stated in the proof of Lemma \ref{lem:cmpc_q_greater_1}, $\Vert x(k+1) - \hat{x}^*(1 \vert k) \Vert \leq r_0(w(k))$, we obtain
\begin{equation*}
	\begin{split}
		&V_{N_p}^* \left( x(k+1),\theta(k+1) \right) - V_{N_p}^* \left( x(k),\theta(k) \right) \leq - \theta(k) \alpha_\ell \left( \Vert x(k) \Vert \right)\\ &+ \xi \sigma_\Gamma \left( r_0 \left( w(k) \right) \right) + \epsilon N_p \bar{\ell}.
	\end{split}
\end{equation*}
We now exploit a $\mathcal{K}$ function $\lambda_2(\cdot)$ such that $\lambda_2(\Vert w \Vert) \geq \xi \sigma_\Gamma \left( r_0 \left( w \right) \right)$ to obtain the final inequality:
\begin{equation*}
	\begin{split}
		&V_{N_p}^* \left( x(k+1),\theta(k+1) \right) - V_{N_p}^* \left( x(k),\theta(k) \right) \leq -\theta(k) \alpha_\ell \left( \Vert x(k) \Vert \right)\\ &+ \lambda_2 \left( \Vert w(k) \Vert \right) + \epsilon N_p \bar{\ell}.
	\end{split}
\end{equation*}

\subsection{Proof of Theorem \ref{thm:cmpc}}
The proof is divided in two parts; we first prove recursive feasibility and then prove input-to-state practical stability.

\textbf{Recursive feasibility:} Lemmas \ref{lem:cmpc_q_greater_1} and \ref{lem:cmpc_q_equal_1}, prove recursive feasibility for all $q^*(k) \in [1,N_p]$. 

\textbf{Input-to-state stability:} Lemma \ref{lem:cmpc_V_{N_p}_star_ineq} proves that $$V_{N_p}^*(x(k),\theta(k)) \leq \xi \Gamma(x(k)) \leq \xi \alpha_\Gamma(\Vert x \Vert) + \epsilon N_p \bar{\ell},$$and, due to the structure of the cost function, we have that $$V_{N_p}^*(x(k),\theta(k)) \geq \theta(k) \ell(x,u) \geq \theta(k) \alpha_\ell(\Vert x(k) \Vert) \geq \epsilon \alpha_\ell(\Vert x(k) \Vert),$$as $\theta(k) \geq \epsilon$.

In the following, we prove that $V_{N_p}^*(x(k),\theta(k))$ has the nominal descent property of typical ISpS-Lyapunov functions, and that the possible rises in its value due to the uncertainty $w(k)$ are bounded. To this end, note that, starting from a feasible initial condition $x(0)$ and imposing $\theta(0) = \max \{ \epsilon, \nu \Gamma \left( x(0) \right) \}$, employing the update rule $\theta(k) = f_\theta \left( x(k),\theta(k-1) \right)$ and setting $\xi \geq 2N_p\bar{\ell}/(1-\gamma)$ ensure that, under Assumptions \ref{ass:cmpc_contractivity}-\ref{ass:cmpc_rcis}, all the conditions of Lemmas \ref{lem:cmpc_q_greater_1} and \ref{lem:cmpc_q_equal_1} are met. 

Lemma \ref{lem:cmpc_q_greater_1} states that, if $q^*(k) > 1$, then the following inequality holds:
\begin{equation*}
	\begin{split}
		&V_{N_p} \left( x(k+1),\theta(k+1),\mathbf{u}^+(k+1),q^+(k+1) \right) - V_{N_p}^* \left( x(k),\theta(k) \right)\\ &\leq -\theta(k) \alpha_\ell \left( \Vert x(k) \Vert \right) + \left( \theta(k)+\xi \right) \lambda_1 \left( \Vert w(k) \Vert \right),
	\end{split}
\end{equation*}
where $q^+(k+1) \coloneqq q^*(k) - 1$, $\mathbf{u}^+(k+1) \coloneqq \big\{ u^*(1 \vert k),\hdots,u^*(q^*(k)-1 \vert k) \big\}$ and $\lambda_1(\cdot)$ is a $\mathcal{K}$ function.

Because of the suboptimality of $V_{N_p} \big( x(k+1),\theta(k+1),\mathbf{u}^+(k+1),q^+(k+1) \big)$, it also holds true that
\begin{equation}
	\label{eq:cmpc_proof_iss_1}
	\begin{split}
		&V_{N_p}^* \left( x(k+1),\theta(k+1)\right) - V_{N_p}^* \left( x(k),\theta(k) \right)\\ &\leq -\theta(k) \alpha_\ell \left( \Vert x(k) \Vert \right) + \left( \theta(k)+\xi \right) \lambda_1 \left( \Vert w(k) \Vert \right).
	\end{split}
\end{equation}

Lemma \ref{lem:cmpc_q_equal_1} states that, if $q^*(k) = 1$, then
\begin{equation}
	\label{eq:cmpc_proof_iss_2}
	\begin{split}
		&V_{N_p}^* \left( x(k+1),\theta(k+1)\right) - V_{N_p}^* \left( x(k),\theta(k) \right)\\ &\leq -\theta(k) \alpha_\ell \left( \Vert x(k) \Vert \right) + \lambda_2 \left( \Vert w(k) \Vert \right) + \epsilon N_p \bar{\ell}.
	\end{split}
\end{equation}

In order to combine \eqref{eq:cmpc_proof_iss_1} and \eqref{eq:cmpc_proof_iss_2}, we make the following consideration.
Given that $\theta(0) = \max \{ \epsilon,\Gamma(x(0)) \}$ and $\theta(k) = f_\theta (x(k),\theta(k-1))$, we have that $\theta(k) \leq \theta(0)$. For this reason, 
\begin{equation*}
	\left( \theta(k)+\xi \right) \lambda_1 \left( \Vert w(k) \Vert \right) \leq \left( \theta(0)+\xi \right) \lambda_1 \left( \Vert w(k) \Vert \right).
\end{equation*}
We now employ a $\mathcal{K}$ function $\lambda(\cdot)$ such that $\lambda(\Vert w \Vert) \geq \left( \theta(0)+\xi \right) \cdot\lambda_1(\Vert w \Vert)$ and $\lambda(\Vert w \Vert) \allowbreak \geq \lambda_2(\Vert w \Vert)$, so as to obtain that, for $q^*(k) \geq 1$,
\begin{equation}
	\label{eq:cmpc_iss_lyapunov}
	\begin{split}
		&V_{N_p}^* \left( x(k+1),\theta(k+1)\right) - V_{N_p}^* \left( x(k),\theta(k) \right)\\ &\leq -\theta(k) \alpha_\ell \left( \Vert x(k) \Vert \right) + \lambda \left( \Vert w(k) \Vert \right) + \epsilon N_p \bar{\ell}.
	\end{split}
\end{equation}
Given that $\theta(k) > 0$ for all $k \geq 0$, \eqref{eq:cmpc_iss_lyapunov}, together with the lower and upper bounds of $V_{N_p}^*(x(k),\theta(k))$, proves that $V_{N_p}^*(x(k),\allowbreak \theta(k))$ is a time-varying ISpS-Lyapunov function, whose variability over time is given by $\theta(k)$.
The existence of a time-varying ISpS-Lyapunov function allows us to state that the equilibrium $x=0$ is ISpS for the closed-loop system.

\subsection{Proof of Theorem \ref{thm:cmpc_2op}}
At each decision instant $k$, the first stage of the algorithm computes the prediction horizon $q^*(k)$ that corresponds to the maximum contraction instant, and the control sequence that leads to such contraction remains available for the second stage. For this reason, every candidate sequence that gives the maximum contraction in the two-stage formulation also verifies all the conditions of Theorem \ref{thm:cmpc}, which implies that the equilibrium $x = 0$ is ISpS for the closed-loop system also under the two-stage formulation of our robust contraction-based MPC.

\section{Algorithms}
\label{sec:algorithms}

\subsection{Computation of the sequences $\{\mathcal{F}(j)\}_{j \geq 0}$ and $\{\mathcal{R}(j)\}_{j \geq 0}$ for component-wise Lipschitz continuous systems}
\label{subsec:F_R_Lipschitz}
In this section, we present a method to compute the sequences $\left\{ \mathcal{F}(j) \right\}_{j \geq 0}$ and $\left\{ \mathcal{R}(j) \right\}_{j \geq 0}$ for component-wise Lipschitz continuous systems, namely systems such that there exist non-negative constants $L_{x,ia}$, $L_{u,ib}$, and $L_{w,ic}$ such that
\begin{equation*}
	\begin{split}
		&\left\vert f_i(x,u,w) - f_i(\check{x},\check{u},\check{w}) \right\vert \\ &\leq \sum_{a = 1}^n L_{x,ia}\vert x_a - \check{x}_a \vert + \sum_{b = 1}^m L_{u,ib} \vert u_b - \check{u}_b \vert \\ &+ \sum_{c = 1}^r L_{w,ic} \vert w_c - \check{w}_c \vert.
	\end{split}
\end{equation*}
In the component-wise Lipschitz case, the previously introduced functions $c_{i,j}(w)$ can be redefined as follows.
For all $i = 1,\hdots,n$, define $c_{i,0}(w) \coloneqq \sum_{c = 1}^r L_{w,ic} \cdot \vert w_c \vert$. Then, for $j > 0$, $c_{i,j}(w) \coloneqq \sum_{a = 1}^n L_{x,ia} \cdot c_{a,j-1}(w)$.

Since $\left\{\mathcal{F}(j)\right\}_{j \geq 0}$ and $\left\{\mathcal{R}(j)\right\}_{j \geq 0}$ have to take into account all the possible realizations of $w \in \mathcal{W}$, we consider the worst case, i.e. a realization $\mathfrak{w}$ such that $\vert \mathfrak{w}_i \vert = \bar{w}_i$, for all $i \in [1,r]$. Therefore, for all $j \geq 0$, $\mathcal{F}(j)$ can be obtained as
\begin{equation*}
	\mathcal{F}(j) = \left\{ x \in \mathbb{R}^n: \; \vert x_i \vert \leq c_{i,j}(\mathfrak{w}), \; \forall i \in [1,n] \right\}.
\end{equation*} 
Now we introduce the constants $d_{i,j} \geq 0$, which are defined for all $i \in [1,n]$. In detail, we impose $d_{i,0} = 0$ and define $d_{i,j}$ for $j > 0$ as $d_{i,j} \coloneqq \sum_{k = 0}^{j-1} c_{i,k}(\mathfrak{w})$. These constants can be used to compute $\mathcal{R}(j)$ as
\begin{equation*}
	\mathcal{R}(j) = \left\{ x \in \mathbb{R}^n: \; \vert x_i \vert \leq d_{i,j}, \; \forall i \in [1,n] \right\}.
\end{equation*} 
Following similar arguments as in \cite{Limon2010_robust_NMPC} and \cite{Manzano2019_Choki}, it can be proved that the sequences $\left\{ \mathcal{F}(j) \right\}_{j \geq 0}$ and $\left\{ \mathcal{R}(j) \right\}_{j \geq 0}$ verify Assumptions \ref{ass:cmpc_F_sequence} and \ref{ass:cmpc_R_sequence}.

\subsection{Computation of the terminal ingredients for nonlinear systems}
\label{subsec:algorithm_terminal_ingredients}
In the following, we assume to deal with a nonlinear unperturbed system described by the state equation $x(k+1) = f \left( x,u \right)$ and that is subject to the following state and input constraints:
\begin{equation*}
	x(k) \in \mathcal{X}, \; u(k) \in \mathcal{U}, \quad \forall k \geq 0.
\end{equation*}
Without loss of generality, we consider the problem of controlling the system state to the origin.

In standard MPC formulations, recursive feasibility and closed-loop stability are ensured by imposing that the state predicted at the end of the prediction horizon belongs to an invariant set $\mathcal{X}_f \subseteq \mathcal{X}$, where a local feasible and stabilizing control law $u = \kappa_f(x)$ is supposed to exist \cite{rawlings2017model}. Hereafter, the dynamics of the system under $\kappa_f(x)$ is referred to as $x(k+1) = f_{cl}(x) \coloneqq f \left( x,\kappa_f(x) \right)$.
Moreover, a terminal cost function $V_f(x)$ is added to the cost function of the optimization problem to be solved, which must be positive definite and such that
\begin{equation}
	\label{eq:terminal_conditions}
	V_f \left( f_{cl}(x) \right) - V_f(x) \leq -\ell(x,\kappa_f(x)), \; \forall x \in \mathcal{X}_f,
\end{equation}
where $\ell(x,u)$ is the stage cost function employed in the MPC.

As often done in the literature, we assume to have a quadratic stage cost function $\ell(x,u) = x^\top Q x + u^\top R u$, to employ a quadratic terminal cost function $V_f(x) = x^\top P x$, with $P$ positive definite, and a local linear controller $\kappa_f(x) = Kx$. We also assume to build the terminal set $\mathcal{X}_f$ as a sublevel set of the terminal cost function $V_f(x)$, i.e. $\mathcal{X}_f \coloneqq \{x: \; x^\top P x \leq \beta\}$, for some $\beta > 0$. 

Assuming to have a controllable couple $(A,B)$ given by the linearization of the system at the origin, and defining the linearization error as $e(x) = f_{cl}(x) - A_Kx$, where $A_K = A+BK$, the condition \eqref{eq:terminal_conditions} rewrites as
\begin{equation*}
	\left( A_Kx + e(x) \right)^\top P \left( A_Kx + e(x) \right) - x^\top P x \leq -x^\top \left( Q + K^\top P K \right) x,
\end{equation*}
or, equivalently,
\begin{equation}
	\label{eq:terminal_conditions_2}
	x^\top \left( A_K^\top P A_K - P + Q + K^\top P K \right)x + e(x)^\top P e(x) + 2x^\top A_K^\top P e(x) \leq 0,
\end{equation}
which must hold true for all $x \in \mathcal{X}_f$.

Assuming that the contribution of the linearization error is bounded, as done in \cite{lazar2018terminal}, i.e.
\begin{equation}
	\label{eq:linearization_error_bound}
	e(x)^\top P e(x) + 2x^\top A_K^\top P e(x) \leq \kappa x^\top P x, \; \forall x \in \mathcal{X}_f,
\end{equation}
for some $\kappa \in (0,1)$, condition \eqref{eq:terminal_conditions_2} can be replaced by the matrix inequality
\begin{equation}
	\label{eq:terminal_conditions_3}
	A_K^\top P A_K - (1-\kappa)P + Q + K^\top P K \leq 0.
\end{equation}
Finally, by using the Schur's complement and the substitution of variables $S = P^{-1}$ and $O = KP^{-1}$, \eqref{eq:terminal_conditions_3} can be rewritten as an LMI:
\begin{equation*}
	\begin{bmatrix}
		(1 - \kappa)S & (AS+BO)^\top & S & O^\top \\
		AS+BO & S & 0 & 0 \\
		S & 0 & Q^{-1} & 0 \\
		O & 0 & 0 & R^{-1}
	\end{bmatrix} \geq 0.
\end{equation*}



\normalsize
\bibliography{references}


\end{document}